\documentclass[twocolumn,times]{aastex62}
\usepackage{amsmath,amstext}
\usepackage{comment}
\usepackage{bm}

\shorttitle{The radio emission of AT2018cqh}
\newcommand{\erg}{${\rm erg \ s^{-1}}$ }

\def\ltsima{$\; \buildrel < \over \sim \;$}
\def\simlt{\lower.5ex\hbox{\ltsima}}
\def\gtsima{$\; \buildrel > \over \sim \;$}
\def\simgt{\lower.5ex\hbox{\gtsima}}

\newcommand{\srcs}{{\rm\,AT2018cqh}}
\newcommand{\src}{{\rm\,AT2018cqh }}

\begin{document}

\title{
%Onset of delayed radio flares from a tidal disruption event in the center of a dwarf galaxy %hosting an active nucleus
%Delayed radio flares from an optical and X-ray bright tidal disruption event in the center of a dwarf galaxy 
% Delayed and fast rising radio flares from an optical and X-ray detected tidal disruption event in the center of a dwarf galaxy 

% Outflow–cloud interaction as nature of peculiar radio emission of the tidal disruption event AT2018cqh
Outflow–cloud interaction as the possible origin of the peculiar radio emission in the tidal disruption event AT2018cqh

%Delayed radio flares from an optical and X-ray detected tidal disruption event  by a candidate intermediate-mass black hole
}
\correspondingauthor{Xinwen~Shu \& Guobin~Mou} 
\email{xwshu@ahnu.edu.cn; gbmou@njnu.edu.cn}

%%\author{Fabao~Zhang }
%\affil{Department of Physics, Anhui Normal University, Wuhu, Anhui, 241002, China}

\author{Lei Yang}
\affil{Department of Physics, Anhui Normal University, Wuhu, Anhui, 241002, China}

\author{Xinwen Shu}
\affil{Department of Physics, Anhui Normal University, Wuhu, Anhui, 241002, China}

\author{Guobin~Mou}
\affil{Department of Physics and Institute of Theoretical Physics, Nanjing Normal University, Nanjing 210023, China}

\author{Yongquan~Xue}
\affil{Department of Astronomy, University of Science and Technology of China, Hefei, Anhui 230026, China}
\affil{School of Astronomy and Space Science, University of Science and Technology of China, Hefei, Anhui, 230026, China}

\author{Luming Sun}
\affil{Department of Physics, Anhui Normal University, Wuhu, Anhui, 241002, China} 

\author{Fabao Zhang}
%%\author{Fabao~Zhang }
\affil{Department of Physics, Anhui Normal University, Wuhu, Anhui, 241002, China}

\author{Zhumao Zhang}
%\author{Fabao~Zhang }
\affil{Department of Physics, Anhui Normal University, Wuhu, Anhui, 241002, China} 

\author{Yibo Wang}
\affil{Department of Astronomy, University of Science and Technology of China, Hefei, Anhui 230026, China}

\author{Tao Wu}
\affil{Department of Physics, Anhui Normal University, Wuhu, Anhui, 241002, China} 

\author{Ning Jiang}
\affil{Department of Astronomy, University of Science and Technology of China, Hefei, Anhui 230026, China}
\affil{School of Astronomy and Space Science, University of Science and Technology of China, Hefei, Anhui, 230026, China}

\author{Hucheng Ding}
\affil{Department of Physics, Anhui Normal University, Wuhu, Anhui, 241002, China} 

\author{Tinggui Wang}
\affil{Department of Astronomy, University of Science and Technology of China, Hefei, Anhui 230026, China}
\affil{School of Astronomy and Space Science, University of Science and Technology of China, Hefei, Anhui, 230026, China}

%\author{Xinwen Shu}
%\affil{Department of Physics, Anhui Normal University, Wuhu, Anhui, 241002, China} %xwshu@ahnu.edu.cn}
%\author[0000-0002-7020-4290]{Xinwen~Shu}
%\affil{Department of Physics, Anhui Normal University, Wuhu, Anhui, 241002, China}

%\author{Lei Yang}
%\affil{Department of Physics, Anhui Normal University, Wuhu, Anhui, 241002, China}

%\affil{Department of Physics, Anhui Normal University, Wuhu, Anhui, 241002, China; 
%ahnuyl@ahnu.edu.cn}

%\author{Luming Sun}
%\author{Fabao~Zhang }
%\affil{Department of Physics, Anhui Normal University, Wuhu, Anhui, 241002, China} 
%ahnuyl@ahnu.edu.cn}

%\author{Zhumao Zhang}
%\author{Fabao~Zhang }
%\affil{Department of Physics, Anhui Normal University, Wuhu, Anhui, 241002, China} 
%ahnuyl@ahnu.edu.cn}

%\author{Yibo Wang}
%\affil{Department of Astronomy, University of Science and Technology of China, Hefei, Anhui 230026, China}

%\author{Guobin~Mou}
%\affil{School of Physics and Technology, Wuhan University, Wuhan 430072, China}
%\affil{WHU-NAOC Joint Center for Astronomy, Wuhan University, Wuhan 430072, China}
%\author{Di~Li}
%\affil{National Astronomical Observatories, Chinese Academy of Sciences, Beijing 100012, China}

%\author{Xue-Guang Zhang}
%\affil{Guangxi Key Laboratory for Relativistic Astrophysics, School of Physical Science and Technology, GuangXi University, Nanning, 530004, China}

%\author{Tianyao~Zhou }
%\affil{Department of Physics, Anhui Normal University, Wuhu, Anhui, 241002, China}

%\author{Fangkun~Peng }
%\affil{Department of Physics, Anhui Normal University, Wuhu, Anhui, 241002, China}

\begin{abstract}

AT2018cqh is a unique optical tidal disruption event (TDE) discovered in a dwarf galaxy exhibiting delayed X-ray and radio flares. 
%Previous research has found that the evolution of delayed radio emission is very peculiar. Not only the lightcurve have a three-stage trend, but also the SED exhibits a gradual shift to a higher peak flux density and frequency. 
We present the results from high-resolution VLBA and e-MERLIN radio observations of AT2018cqh extending to $\delta$t $\sim$ 2250 days post discovery, 
%as well as that from the ongoing ASKAP observing compaigns. The VLBA and e-MERLIN observations 
which reveal a compact radio emission, unresolved at a scale of $\simlt$0.13 pc at 7.6 GHz, 
%using the deconvolved beam size of VLBA 7.6GHz observation, 
with a high brightness temperature of $T_b$ $\simgt$ 4.03 $\times$ 10$^{9}$ K.
The radio spectral energy distribution (SED) is found to gradually shift towards a higher peak flux density and frequency over a period of 
$\sim$1000 days. 
An equipartition analysis suggests that there is a little change in the radio emitting region over this period, 
%a period of {\bf $\sim 1000$} days, 
while the electron density increases by a factor of 3. 
The radio light curve at 0.89 GHz continues to rise, with a bump feature lasting for 240 days.   
%The continuous archive data from VAST and RACS show that the evolution of the lightcurve has become flat, accompanied by flux oscillations. 
These properties are in contrast to the predictions of standard shockwave model from a diffuse circumnuclear medium, but could be explained if dense clouds exist in the circumnuclear environment. 
The latter scenario is supported by our hydrodynamic simulations of the interaction of TDE outflow with a cloud, which can reproduce the temporal evolution in the radio SED. 
This work highlights the importance of the outflow-cloud interaction in explaining the delayed, 
fast-rising radio emission observed in some TDEs, 
especially those occurring in galaxies with pre-existing AGN activity.  

%Our results demonstrate that \src is the best case known so far whose radio emission likely originates from the outflow-cloud interaction. 
%as demonstrated in the simulations 
%outflow-CNM model, whereas the rapid rise in radio flux, the constant size of the emission region obtained through equipartition analysis, and the increasing ambient density can all be well interpreted by the interaction between outflow and cloud.
% These observational properties are inconsistent with the model predictions of the outflow-CNM interaction, but can be naturally explained by the model of the outflow-cloud interaction. 
%We performed simulated the data using ZEUS-3D and found that the results were consistent with the data well, which indicates that the radio emission in AT2018cqh is likely originates from the outflow-cloud interaction.
%Our results %presents a  
%highlight that the clumpy gas clouds 
%the critical role  

\end{abstract}


\section{Introduction} \label{sec:intro}

When a star passes too close to a supermassive black hole (SMBH), it can be squeezed and torn apart once the tidal force of the SMBH exceeds the star's self-gravity \citep{Rees1988}, which is called a tidal disruption event (TDE). 
% Such tidal disruption events (TDEs) can generate luminous flares typically peaking in X-rays and ultraviolet as 
Approximately half of the debris remains in orbits bound to the black hole, while other parts are flung out on hyperbolic orbits \citep{Rees1988}. The bound stellar debris will circularize and form an accretion disk,  %accreting at sub-Eddington 
generating luminous flares typically peaking in X-rays and ultraviolet \citep{Saxton..2020, Gezari2021}. 
%providing a direct way to probe SMBHs in otherwise quiescent galaxies \citep{Mockler..2019}.
TDEs enable a direct %measurements of accretion events 
probe to the transient accretion 
onto SMBHs, as well as the subsequent launch of jet and outflow. 
When a jet/outflow propagates outward, it will interact with the diffuse circumnuclear medium (CNM). The shock process in CNM can accelerate eletrons in magnetic field, producing a bright flare of radio synchrotron emission  \citep{DeColle..2012, Lu..2018, Alexander..2020, Hu..2025}. 
Therefore, radio observations of TDEs are not only essential to infer the properties of jet and outflow, but also to illuminate the environment around otherwise dormant SMBHs \citep{Alexander..2020}.

%Among TDEs that have radio detections, a few 
%Although very few TDEs show evidence of powerful relativistic jets, 
%including Swift J1644+57, Swift J1112, Swift J2058, and AT2022cmc 
%\citep{Eftekhari..2018, Brown..2015, Pasham..2015, Andreoni..2022}, 
%they are readily detected at radio frequencies since discovery. 
%In addition the jetted ones, radio follow-up observations of TDEs 
%discovered in other bands have resulted in very few conclusive detections of radio emission (Bower et al. 2013; vanVelzen et al. 2013). 
%and possibly relativistic jets  
% Another classic radio TDE shows evidence for only a sub-relativistic
%Among TDEs that have radio detections, a few show evidence of 
So far only four TDEs have been securely identified to have powerful on-axis relativistic jets  
%such as Swift J1644+57, Swift J1112, Swift J2058, and AT2022cmc 
\citep{Burrows..2011, Brown..2015, Pasham..2015, Andreoni..2022}, 
which were readily detected at radio frequencies since discovery. 
Most radio-emitting TDEs could be related to off-axis jets \citep{Lei..2016, Mattila..2018, Sfaradi..2024}, sub-relativistic \citep{Cendes..2022, Cendes..2025}, or non-relativistic wide-angle outflows \citep[e.g.,][]{Alexander..2016, Alexander..2017, Cendes..2021b, Cendes..2022, Goodwin..2022, Goodwin..2023a, Goodwin..2023b, Goodwin..2024}, 
% Most radio-emitting TDEs could be related to off-axis jets, sub-relativistic, or non-relativistic wide-angle outflows \citep[e.g.,][]{Lei..2016, Alexander..2016, Alexander..2017, Mattila..2018, Cendes..2022, Goodwin..2022, Goodwin..2023a, Goodwin..2023b, Goodwin..2024}, 
with radio luminosities about two orders of magnitude lower than on-axis jetted TDEs. 
% (Alexander et al. 2016; Alexander et al. 2017; Cendes et al. 2021, 2022; Goodwin et al. 2022, 2023a, 2023b, 2024a, 2024b). 
% While the energetic radio emission observed from relativistic TDEs is consistently explained by a relativistic jet launched from the SMBH
%While relativistic jets from SMBHs consistently explain the energetic radio emission in relativistic TDEs \citep[e.g.][]{Bloom..2011, Cenko..2012, Pasham..2023}, the mechanism behind the lower energy radio emission in non-relativistic TDEs remains debate. 
%Possible scenarios include an outflow produced by disc winds or a sub-relativistic jet \citep[e.g.][]{Alexander..2016, vanVelzen..2016, Pasham..2018}, collision-induced outflow from stream-stream collisions of the stellar debris \citep{Lu..2020}, and outflow from unbound tidal debris \citep{Krolik..2016, Yalinewich..2019}.
Recently, \cite{Cendes..2024} present the results from radio follow-up observations of 23 optical TDEs and find that $\sim 40\%$ of them exhibit radio brightening hundreds to thousands of days after the discovery, indicating the ubiquitous late radio emission from TDEs though the origin remains unclear.   %\cite{Cendes..2024} propose delayed disk formation as the origin of the late-time outflow.
%which could be explained by the delayed launching of outflow and its interaction with CNM.  
% Surprisingly, recent radio observations have revealed delayed but rapidly rising radio emission in some TDEs.
% Recent TDE radio observations reveal flares only seen at late time ($10^{2\sim3}$ days after TDE discovery) and steep rise in light-curve, such as ASASSN-15oi, AT2018hyz and AT2018cqh 
%Recent radio observations of TDEs (e.g., ASASSN-15oi, AT2018hyz, AT2018cqh) reveal late-time flares ($10^{2\sim3}$ days post-discovery) with 
More interestingly, a few TDEs display %Monitoring the flux evolution of the delayed radio emission reveals 
steeply rising light curves in the delayed radio emission \citep{Horesh..2021a, Cendes..2022, Zhang..2024, Sfaradi2025},
%These flares exhibit a flux density jump from nondetection to detection, 
requiring a temporal power-law index steeper than $t^4$—a behavior inconsistent with the predictions of standard model that a single outflow (either relativistic or sub-relativistic) is launched into the CNM promptly following the TDE. 
%outflow-CNM models.

Several scenarios have been proposed to explain these late-time and steeply rising radio flares. 
% \cite{Cendes..2024} suggest that delayed outflow from delayed disk formation is a preferred explanation. 
% \cite{Cendes..2024} propose delayed disk formation as the origin of the late-time outflow.
% However, early X-ray detection indicates that the accretion disk is already present early on \citep{Guolo..2024}.
% Alternatively, \cite{Matsumoto..2023} propose deceleration of an off-axis jet launched at the time of disruption. However, the initial steep flux rise and the complex spectral and temporal evolution of some TDEs are inconsistent with the spectral index evolution predicted by deceleration jet(Horesh et al. 2021, Zhang et al. 2024).
\cite{Matsumoto..2023} suggest that a steep rise in the radio light curve can be obtained if a jet 
is observed off-axis that became visible at late times \citep{Sfaradi..2024},  
%However, such an off-axis jet model 
but it cannot account for the big jump in flux density from non-detection to detection \citep{Horesh..2021a}. 
%Yet, the rapid initial flux rise and complex spectral evolution in some TDEs \citep{Horesh..2021a, Zhang..2024} conflict with predictions from jet deceleration models.
\cite{Teboul..2023} propose a scenario invoking the break out of a misaligned precessing jet that is initially choked by the disk wind \citep[also see,][]{Lu..2024}, indicating a radio-emitting process that occurs at late times.  %but later breaks out when the disk eventually aligns with the BH spin axis. 
%Further, \cite{Matsumoto..2024} find that a CNM density profile that initially drops with radius in a power law but is followed by a constant outside the Bondi radius. However, it is difficult for this model to explain a rise steeper than $\propto t^3$, like those in AT2018hyz and ASASSN-15oi, unless the external density profile rises again. 
%\cite{Lei..2024} studied 
It is also possible that there are multiple outflows launched at different times relative to the discovery of TDE, which might be responsible for the fast radio brightening at a later time \citep{Sfaradi2025}. 
While the collision between unbound stellar debris and dense gas from the inner edge of a dusty torus (which is observed to exist in Active Galactic Nuclei, AGN) 
can explain the fast rising radio light curve {\citep{Lei..2024}}, the predicted radio luminosity appears too low 
to reconcile with the data observed in the TDE ASASSN-15oi and AT2018hyz. 
In addition, by simulating the TDE evolution in the presence of an AGN disk, \citet{Chan..2019} suggests 
that part of outgoing debris stream will shock with surrounding gas, generating a bright and possibly delayed   
radio synchrotron emission, though they did not perform detailed calculations of the radio light curve.

Alternatively, if the CNM is clumpy and filled with dense clouds, 
the expanding TDE outflow will sweep across them, producing bright radio emission in the bow shock \citep{Mou&Wang2021, Mou..2022, Bu..2023a}, which can naturally explain the delayed, fast rising light curve due to the sharp edge of cloud, 
as well as radio rebrightenings observed in some TDEs \citep{Zhuang..2025}. 
%In addition to the TDEs discovered by optical surveys, 
While many TDEs are identified in normal galaxies \citep[e.g.,][]{Hammerstein..2023}, there is a growing number of TDEs that have been discovered in galaxies with pre-existing AGN activity %to occur in AGNs
%, such as GSN 069, AT2021aeuk, IRAS F01004-2237, and AT2019aalc 
\citep[e.g.,][]{Blanchard..2017, Shu..2018, Zhang2022, SunLM..2024, Sun..2025, Veres..2024}. 
%The material environment around black holes in AGN may differ from that around quiescent black holes. 
% There is consensus that exists a so-called broad-line region (BLR) composed of large amounts of clouds surrounding the central black hole in AGN.}
%It is generally agreed that a so-called broad-line region (BLR), composed of numerous clouds, surrounds the central black hole in AGN.
For the latter case, the existence of broad-line region clouds at a sub-pc scale \citep[e.g.,][]{Armijos2022} provides the necessary condition for the interaction between TDE
 outflow and clouds, allowing for investigating in detail on the physical process and 
 the resulting radio emission properties. %scenario. 
% perhaps producing radio emission with properties different from 
%those in CNM. %occurring in normal galaxies.   %that might be different  

%in a gas-rich circumnuclear environment, discrete clouds might exist
%{\bf cloud?}
%\cite{Mou..2022} studied the outflow–cloud interaction and showed that it can generate considerable radio emission months or years later after the TDE.
% and can explain the temporal evolution of the peak frequencies in AT2019dsg, ASASSN-14li, and CSS161010.
%Based on numerical simulations of the disk wind launching, \cite{Bu..2023a} further studied this model.
% Besides the delayed radio flare, the outflow–cloud interaction may have multi-messenger signatures including neutrinos, gamma rays, and X-rays \citep{Mou&Wang2021, Mou..2021, Wu..2022, Chen&Wang2023}. 
%Furthermore, \cite{Zhuang..2025} developed the outflow-cloud model for the TDEs with late-time flares and steep rising light curves, found this model can generating bright radio flares years after TDE with a luminosity of 10$^{39} erg s^{-1}$ and can explain the large delay, the sharpness of the rise, and the multiplicity of the late radio flares.

%SRGe J023346.8-010129 
AT2018cqh is a rare faint TDE that occurred in a dwarf galaxy at $z=0.048$ \citep{Bykov..2024} and exhibited optical, X-ray and radio flares, which were likely driven by an intermediate-mass black hole.
A striking feature is its delayed radio emission, characterized by a two-phase fast rising light curve \citep{Zhang..2024}, 
the nature of which is enigmatic. 
% It was first discovered by Gaia on 2018 June 16, and named Gaia18bod. In this paper, we use the name AT 2018cqh as given in the TNS. 
% \cite{Zhang..2024} have analyzed the radio characteristics
%and discussed the nature of late-time radio flare in AT2018cqh 
% and found that its radio evolution can be divided into three phases: an initial steep rise, followed by a plateau, and another steep rise. Additionally, the radio SED moves towards higher peak frequency and peak flux, and the physical parameters calculated using the energy distribution model also increase over time. 
% This is very different from the typical radio evolution observed in ordinary TDEs. 
% In contrast, this is very different from the typical radio evolution observed in ordinary TDEs.
% In this paper, we supplement the latest 0.89 GHz observations from ASKAP and provide an explanation of the nature of radio emission in AT2018cqh within the framework of outflow-cloud interaction, combining high-resolution observations from VLBA and e-MERLIN. The results show that the outflow-cloud interaction can reasonably fit all three epoch SED and radio light curves. 
% Importantly, our study indicates that not all of the radio emission in TDEs can be modeled using the energy distribution model to calculate physical parameters, and a significant part of the radio radiation in TDEs may originate from outflow-cloud interactions. 
%\cite{Zhang..2024} have analyzed the radio light curve and SED of AT2018cqh and found that its evolution significantly differs from typical TDEs. 
In this Letter, we report the results of the high-resolution Very Long Baseline Array (VLBA) and enhanced-Multi-Element Remotely Linked Interferometer Network (e-MERLIN) follow-up observations, as well as the analysis of the newly released data from Australian Square Kilometre Array Pathfinder (ASKAP) observations at low frequency, 
%covering a new period spanning 
extending to $\delta t\sim2510$ days since optical discovery \citep[$\rm MJD=58285$,][]{Zhang..2024}. 
We find a peculiar radio spectral evolution peaking toward a higher frequency and flux density over a period of $\sim$1000 days. 
Moreover, while the light curves continue to rise, there is a bump feature lasting for 240 days, with a fast rise to peak time of only 115 days. 
There properties are distinct from other TDEs, 
%\textbf{Combined with the fact that AT2018cqh resides in a type 2 AGN\citep{Zhang..2024}, where the central black hole was once in an accretion state, the material around its black hole may not be uniformly distributed. Instead, it could contain dense clouds similar to those in the BLR.} This 
and point to the scenario that an outflow is launched 
around the time of optical discovery and interacting with dense clouds at a sub-pc scale. 
This is possible as the host of \src can be spectroscopically classified as a type 2 AGN %\citep{Zhang..2024} 
and the dense clouds are likely coming from a pre-existing broad-line region. 
%in which the broad-line region clouds are present before the TDE. 
%cannot be explained under a standard CNM shockwave model. 
% Our findings reveal that AT2018cqh is compact at milliarcsecond resolution of VLBA.
%Fascinatingly, the quasi-simultaneous SED from VLBA and e-MERLIN observations between Aug 2025 and Sep 2025 reveal that 
% both the peak frequency and peak flux density continue shifting toward higher values.
%not only has the peak frequency shifted toward higher frequencies, but the peak flux density has also shown an increasing trend. Additionally, the flux density of ASKAP 0.89GHz exhibits oscillatory behavior.}
% Using the equipartition method, we find that the size of the emitting region remains nearly constant across three epochs, while the ambient density keeps increasing, which aligns well with the predictions of the outflow-cloud model. 
%Consequently, we performed hydrodynamical simulations of the radio light curve and the SED evolution of AT2018cqh using the ZEUS-3D code, demonstrating that its radio properties can be well explained by the outflow-cloud scenario.
The observations and data reductions are described in Section 2. In Section 3, we present the detailed analysis of radio morphology, flux and SED evolution properties. 
In Section 4, we perform hydrodynamic simulations to explore the origin of the peculiar radio flux and SED evolution in the context of the interaction between outflow and surrounding clouds. % the origins of delayed radio from AT2018cqh is given in Section 4. 
We summarize the results in Section 5. 
%We adopt a cosmology of $\Omega_M$ = 0.3, $\Omega_{\lambda}$ = 0.7, and H$_0$ = 70 km s$^{-1}$ Mpc$^{-1}$ when computing luminosity distance.
Throughout this paper, we assume a cosmology with $H_0$ = 70 km s$^{-1}$ Mpc$^{-1}$, $\Omega_M$ = 0.3, and $\Omega_{\Lambda}$ = 0.7, corresponding to a scale of 0.941 pc milliarcsec$^{-1}$ at the redshift of AT2018cqh.

\section{OBSERVATIONS AND DATA REDUCTION} \label{sec:style}

\begin{deluxetable*}{ccccccc}
  \centering
\tablewidth{0pt}
\tablehead{
\colhead{Observatory} & \colhead{Project} & \colhead{$\nu$} & \colhead{Date} & \colhead{Phase} & \colhead{F$_\nu$} & \colhead{rms}\\
\colhead{} & \colhead{} & \colhead{(GHz)} & \colhead{} & \colhead{(days)} & \colhead{(mJy/beam)} & \colhead{(mJy/beam)}}
\tablecaption{Summary of the radio observations of AT2018cqh \label{tab:table}}
\setlength{\tabcolsep}{3mm}{
\startdata
ASKAP & RACS & 1.66 & 2022 Jan 08 & 1302 & 3.331 $\pm$ 0.080 & 0.183\\
\hline
ASKAP & VAST & 0.89 & 2023 Dec 26 & 2019 & 6.640 $\pm$ 0.160 & 0.252\\
 & RACS & 0.94 & 2024 Jan 03 & 2027 & 8.880 $\pm$ 0.130 & 0.165\\
 & VAST & 0.89 & 2024 Mar 02 & 2086 & 4.240 $\pm$ 0.210 & 0.249\\
 & VAST & 0.89 & 2024 Apr 27 & 2142 & 4.770 $\pm$ 0.150 & 0.298\\
 & VAST & 0.89 & 2024 Jun 21 & 2197 & 6.250 $\pm$ 0.160 & 0.306\\
 & VAST & 0.89 & 2024 Aug 21 & 2258 & 6.960 $\pm$ 0.160 & 0.230\\
 & VAST & 0.89 & 2024 Oct 22 & 2320 & 6.260 $\pm$ 0.150 & 0.226\\
 & VAST & 0.89 & 2024 Dec 21 & 2380 & 9.330 $\pm$ 0.170 & 0.231\\
 & VAST & 0.89 & 2025 Feb 22 & 2443 & 6.510 $\pm$ 0.140 & 0.235\\
 & VAST & 0.89 & 2025 Apr 21 & 2510 & 11.80 $\pm$ 0.250 & 0.267\\
\hline
VLBA    & BS340B & 1.56 & 2024 Aug 04 & 2241 & 10.91 $\pm$ 0.780 & 0.356\\
        & BS340A & 4.74 & 2024 Aug 06 & 2243 & 16.023 $\pm$ 0.056 & 0.076\\
        &        & 7.63 & 2024 Aug 06 & 2243 & 11.574 $\pm$ 0.044 & 0.061\\
\hline
e-MERLIN & CY18012 & 1.51 & 2024 Sep 28 & 2297 & 10.310 $\pm$ 0.380 & 0.220\\
         &         & 5.07 & 2024 Sep 02 & 2270 & 12.210 $\pm$ 0.180 & 0.147\\
\hline
%uGMRT & ddtC411 & 0.75 & 2025 Jan 03 & 2297 & 10.310 $\pm$ 0.380 \\
uGMRT & ddtC411 & 1.26 & 2024 Dec 29 & 2388 & 14.120 $\pm$ 0.380 & 0.075\\
      &         & 0.75 & 2025 Jan 25 & 2415 & 5.250 $\pm$ 0.280 & 0.274\\
      &         & 1.26 & 2025 Jan 26 & 2416 & 20.620 $\pm$ 0.650 & 0.056\\
      &         & 0.75 & 2025 Feb 21 & 2442 & 5.680 $\pm$ 0.350 & 0.496\\
      &         & 1.26 & 2025 Feb 20 & 2441 & 23.120 $\pm$ 0.270 & 0.066\\
      &         & 0.75 & 2025 Mar 24 & 2473 & 4.600 $\pm$ 0.530 & 0.466\\
      &         & 1.26 & 2025 Mar 22 & 2471 & 19.600 $\pm$ 0.260 & 0.070\\
\hline
\enddata}
\label{radio data}
\end{deluxetable*}

% \begin{figure}[htbp!]
%   \begin{center}
% \includegraphics[width=0.50\textwidth]{J0233-0101_LC_obsframe_20240603.pdf}
%  \hspace{0.1cm}
%  \end{center}
%   \vspace{-0.5cm}
% \caption{The light curves of AT 2018cqh in the optical (ZTF r band and Gaia G band), X-ray (0.3–2 keV), and radio (ASKAP 0.89GHz, 1.36GHz and 1.66GHz, VLASS 3GHz, and VLBA 1.56GHz, 4.74GHz, and 7.63GHz). Note that the optical light curves include the flux contributed by the host. It is clear that there is a delayed brightening in the X-rays, while AT 2018cqh has faded in the optical to a quiescent level. The radio emission appears even later and showed an likely oscillating trend at 0.89 GHz. For the nondetections, the corresponding 3σ upper limits on flux density are shown.}
% \label{fig:radiolc}
%  \vspace{0.2cm}
% \end{figure}

\subsection{e-MERLIN}

To investigate possible radio flux and SED evolution of \srcs,  %between the mas scale of the VLBA and the arcsecond scale of ASKAP, 
we performed follow-up radio observations using e-MERLIN. e-MERLIN is a UK-based radio interferometer with a maximum baseline of 217 km and seven dishes spanning 25$\sbond$76 m in diameter. The facility can observe at $L-, C-$, and $K-$band. The e-MERLIN observations of \src were completed on 2024 Sep 28 at $L$ band (central frequency of 1.51 GHz) and on 2024 Sep 02 at $C$ band (central frequency of 5.07 GHz) (project code: CY18012). The bandwidth is 0.52 GHz for $L$-band and 0.51 GHz for $C$-band, respectively. Data calibration was performed with the e-MERLIN Common Astronomy Software Applications (CASA) pipeline \citep{Moldon2021} using standard techniques. 
Bright radio emission components are detected in both bands. 
We used the {\tt IMFIT} task in CASA to fit the radio emission with a two-dimensional 
elliptical Gaussian model to determine the position, integrated flux density, and peak flux density. 
%There is no extended emission 
The radio emission at both bands is unresolved and no extended emission is detected. 
e-MERLIN detects a compact source in the final cleaned image, which has a clean beam size of 0.41 $^{\prime\prime} \times 0.13 ^{\prime\prime}$ at 1.51 GHz, 0.12 $^{\prime\prime} \times 0.04 ^{\prime\prime}$ at 5.07 GHz.
%The compactness of the radio emission is confirmed by the ratios of integrated and peak flux density, 
%which are in the range 0.92--1.35, with a median of 1.09. 
We obtained a peak flux density of 10.31 $\pm$ 0.38 mJy/beam at $L$ band and 12.21 $\pm$ 0.18 mJy/beam at $C$ band, respectively. 
Note the the integrated and peak flux densities are roughly equal to each other at both bands, as expected for a compact radio source. 
Therefore, for consistency, only peak flux densities are used in our following analysis.
The e-MERLIN observational log and flux density measurements are presented in Table \ref{radio data}. 
%Bright radio emission components are detected in both bands, with flux density of 10.31 $\pm$ 0.38 mJy/beam at $L$ band and 12.21 $\pm$ 0.18 mJy/beam at $C$ band, respectively. Observation results are reported in Table \ref{radio data}.

\begin{figure*}[t!]
\centering  
\includegraphics[width=0.97\textwidth]{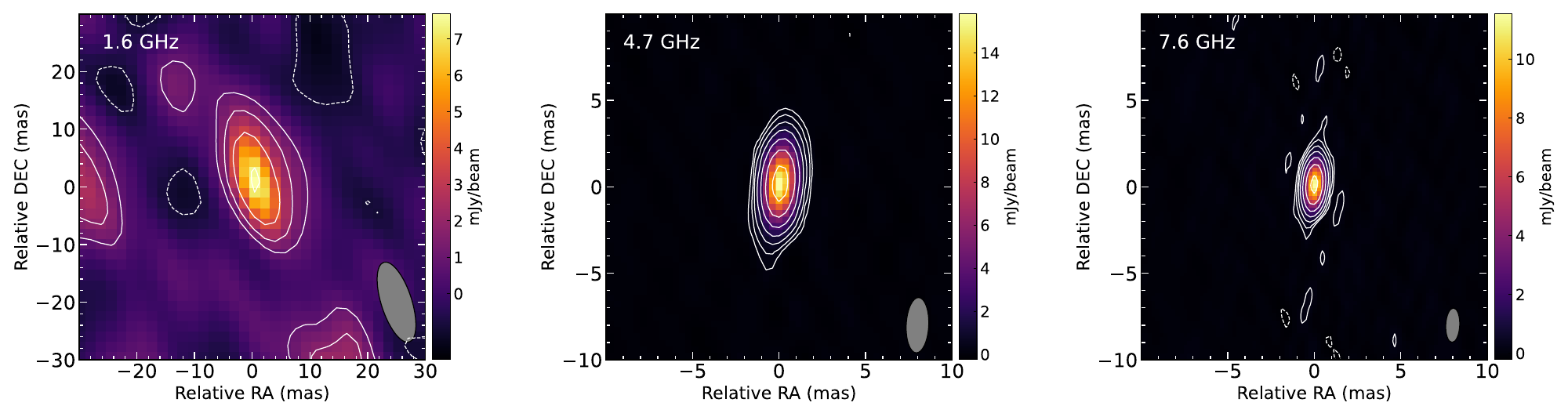}
\caption{Natural-weighted images of VLBA observations of AT 2018cqh at three frequencies. In each image, the bottom right ellipse is the shape of the restoring beam, with sizes of 14.51 $\times$ 5.29 mas for L-band, 3.20 $\times$ 1.30 mas for C-band, and 1.99 $\times$ 0.81 mas for X-band, respectively. The contour levels are 2.5$\sigma \times$(-1, 1, 2, 4, 8) for L-band, and 2.5$\sigma \times$(-1, 1, 2, 4, 8, 16, 32, and 64) for C-band and X-band.
\label{fig: vlba}}
\end{figure*}

\subsection{VLBA}

The VLBA observations of \src were carried out on 2024 Aug 04 at $L$ band (central frequency of 1.57GHz) and on 2024 Aug 06 at $C$ band (central frequencies of 4.74GHz and 7.63GHz) (project code: BS340). Seven antennas were participated in the observations (BR, FD, HN, NL, OV, PT, and SC for $L$ band observation; BR, FD, HN, MK, NL, PT, and SC for $C$ band observation). The observations were performed in the phase-referencing mode. The bright source J0239-0234 ($\sim$420 mJy at 5GHz, about 2.15$\degr$ from the target) was used as the phase-reference calibrator. Phase-reference cycle times were 4.5 minutes, with 3.5 minutes on-source and 1.0 minutes for the phase calibrator. We also inserted several scans of the bright source 3C84 for fringe and bandpass calibration with an integration time for each scan of 4 minutes at $L$ band and 3 minutes at $C$ band, respectively. To check the calibration results, we also inserted several scans of a bright source J0223-0205 ($\sim$120 mJy at 8.7GHz, about 2.85$\degr$ from the target) with an integration time of 2 minutes for each scan. The resulting total on-source time was 2 hours for $L$ band and 4 hours for $C$ band, respectively. To achieve sufficiently high imaging sensitivity, we adopted the observational mode Digital Down-converter System for Roach Digital Backend to use the largest recording rate of 2 Gbps, corresponding to a recording bandwidth of 128 MHz in each of the dual circular polarizations. The data from the VLBA experiment were correlated with the DiFX software correlator \citep{Deller..2011}. We used the NRAO AIPS software to calibrate the amplitudes and phases of the visibility data, following the standard procedure from the AIPS Cookbook\footnote{\url{http://www.aips.nrao.edu/cook.html}}. The calibrated data were imported into the Caltech DIFMAP package \citep{Shepherd1997} for imaging and model fitting. 
We then measured the flux densities at 1.6 GHz, 4.7\,GHz and 7.6\,GHz, %were then measured 
following the procedures described in Section 2.1. 
The results are given in Table 1\footnote{
%The VLBA $L$-band showing a slight discrepancy between the peak flux density (7.25mJy) and int flux density (10.91mJy). However, 
Due to the presence of severe radio frequency interference (RFI) in the VLBA observation at $L$-band, a considerable portion of the data was flagged during calibration process. This may introduce errors in phase calibration, resulting in a relatively high image rms of $\sim$0.4 mJy/beam and larger uncertainty in the peak flux. Therefore, we report the integrated flux density for the VLBA $L$-band observation that is more reliable.}.

\subsection{uGMRT}

AT2018cqh was observed with the upgraded Giant Metrewave Radio Telescope (uGMRT) at band4 (central frequencies of 0.75 GHz) on 2025 Jan 03, 2025 Jan 25, 2025 Feb 21, and 2025 Mar 24, and at band5 (central frequencies of 1.26 GHz) on 2024 Dec 29, 2025 Jan 26, 2025 Feb 20, and 2025 Mar 22 (project code: ddtC411). Flux calibration was conducted using 3C 48, whereas the nearby source 0217+017 was also used to determine the complex gain solutions. 
% We note that the band 5 observation of project ddtC206 cannot be imaged as the short baselines have problems estimating gain solutions for the target. 
The band4 observation performed on 2024 Jan 03 could not be imaged due to the absence of radio interferometric fringes from the phase calibrator.
The data from the uGMRT observations were reduced using CASA (version 5.6.1) following standard procedures and by using a pipeline adapted from the CAsa Pipeline-cum-Toolkit for Upgraded Giant Metrewave Radio Telescope data REduction \citep[CAPTURE; ][]{Kare..2021}. We began our reduction by flagging known bad channels, and the remaining RFI was flagged with the flagdata task using the clip and tfcrop modes. We then ran the task tclean with the options of the MS-MFS \citep[multi-scale multi-frequency synthesis, ][]{Rau..2011} deconvolver, two Taylor terms (nterms=2), and W-Projection \citep{Cornwell..2008} to accurately model the wide bandwidth and the non-coplanar field of view of uGMRT. AT2018cqh was detected in all uGMRT observations as an unresolved source, and the flux density measurements are shown in Table \ref{radio data}.
% While AT2018cqh was not detected in the first and second epoch uGMRT observations, it appeared in the third epoch as an unresolved source with a peak flux of 0.349 ± 0.055 mJy/beam at band 4 and 0.558 ± 0.058 mJy/beam at band 5. All the uGMRT flux density mea- surements are shown in Table 1.

In addition, %the field of \src The Very Large Array Sky Surve \citep[VLASS,][]{Lacy..2020} archive data, 
\src was repeated observed at low frequencies by ASKAP Variables and Slow Transients Survey \citep[VAST;][]{Murphy..2021}, and the Rapid ASKAP Continuum Survey \citep[RACS;][]{McConnell..2020}.  
%up to Apr 21, 2025. 
%have been presented in the article by \cite{Zhang..2024}. We note that the VAST are still continuing to survey the area where AT2018cqh is located. 
%While part of VAST and RACS data have been analyzed in \citet{Zhang..2024}, 
Here we present an uniform analysis of all archival radio data obtained with VAST and RACS, 
specifically, for those that are not given in \citet{Zhang..2024}, %added the latest VAST data 
covering a new period from Dec 26, 2023 to Apr 21, 2025. 
%and the recently released RACS 1.66GHz data which observed in 2022 Jan 08. 
%We determined the source coordinates, integrated and peak flux density by using the same method as in Section 2.2 of \cite{Zhang..2024}. 
Table \ref{radio data} presents the ASKAP observational log and flux density measurements. 
%are presented in Table \ref{radio data}.

\begin{figure}[t!]
  \centering  
  \includegraphics[width=0.45\textwidth]{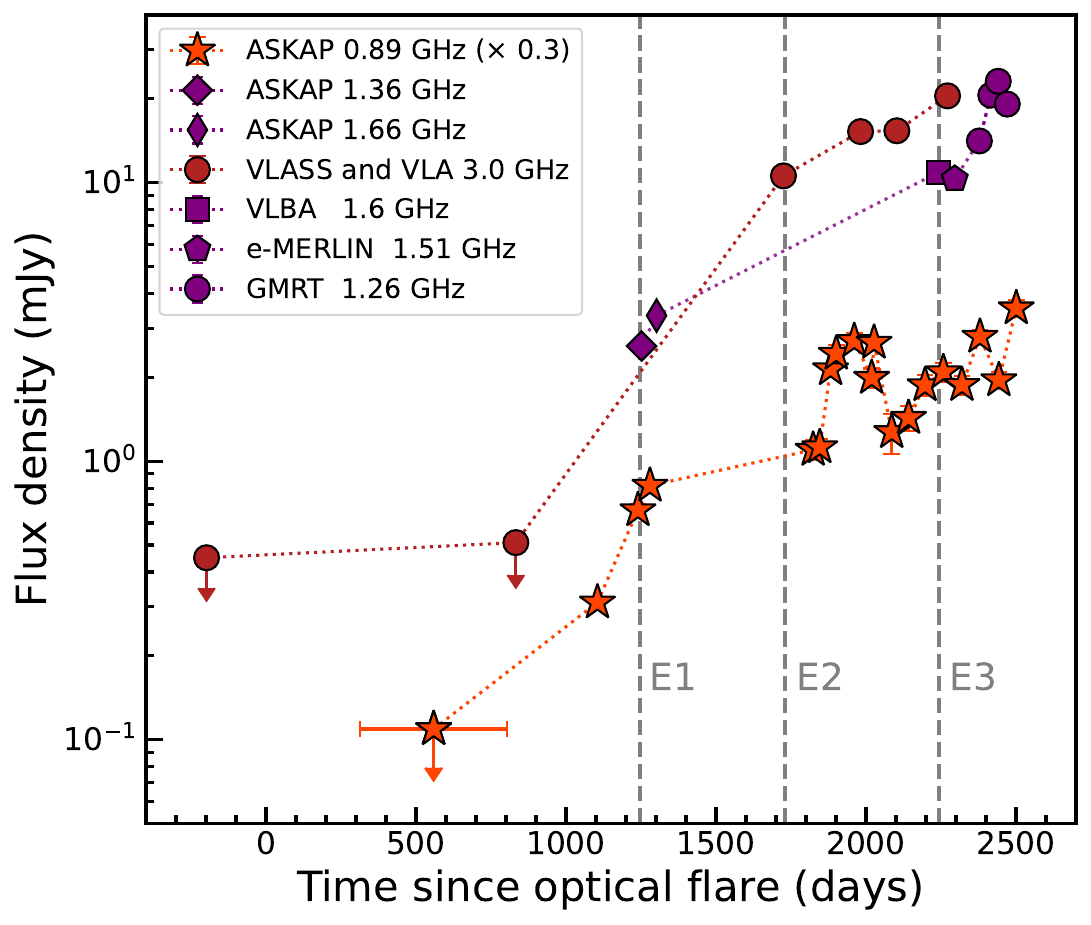}
  \caption{The radio light curves of AT 2018cqh, including the data from the ASKAP observations at 0.89 GHz, 1.36 GHz and 1.66 GHz, the e-MERLIN observations at 1.51 GHz, the VLASS observations at 3 GHz, the VLBA observations at 1.56 GHz, and the VLA observations at 3.0 GHz (Section 3.2). %(project code: 23B-321; PI: Collin Christy). 
  The radio emission appears later relative to the time of optical discovery ($\rm MJD=58285$) and shows a continuous rising in flux that can vary on time-scales down to $\sim$2 months. 
  %brightening with flux variability down to to $\sim$2 months. %possibly oscillating trend at 0.89 GHz. 
  For the nondetections, the corresponding 3$\sigma$ upper limits on flux density are shown. The vertical dashed lines represent the three epochs for which the radio SED can be constrained (Figure \ref{fig:radiosed}), 
  %of the three SED observations, which occurred at 
  corresponding to 1246, 1730, and 2243 days after the optical discovery, respectively.
  }
  \label{fig:radiolc}
\end{figure}

\section{Analysis and Results} 

\subsection{Parsec-scale Radio Morphology}

In Figure \ref{fig: vlba}, we show the VLBA images of AT2018cqh at 1.56, 4.74, and 7.63GHz, respectively. Because the signal-to-noise ratio of the its radio emission is high,
% \srcs' 
we performed one loop phase self-calibration at $C-$ and $X-$band. After several tens of fitting iterations in CASA by using an elliptical Gaussian brightness distribution model, we measured the integrated flux densities of C- and X-band which are 17.95 $\pm$ 0.12 mJy and 12.80 $\pm$ 0.091 mJy, respectively, while the corresponding peak flux densities are 16.02 $\pm$ 0.056 mJy and 11.57 $\pm$ 0.044 mJy. The ratio of the integrated and peak flux densities is 1.12 and 1.10 for C- and X-band, indicating that the radio source might be unresolved 
in the final imaging.
%Besides the central peak, no extended emission was seen above the five?? times of the image noise level. 
%VLBA detects a compact source in the final cleaned image, which has a 
The size of restoring beam\footnote{The restoring beam is an idealized elliptical Gaussian that approximates the main lobe of the dirty beam. It is determined by the UV coverage of the array.} is 14.51 $\times$ 5.29 mas at 1.56 GHz, 3.20 $\times$ 1.30 mas at 4.74 GHz, and 1.99 $\times$ 0.81 mas at 7.63 GHz, respectively.
The deconvolved beam size is 7.66 mas $\times$ 4.64 mas at 1.56 GHz, 0.996 mas $\times$ 0.381 mas at 4.74 GHz, and 0.676 mas $\times$ 0.134 mas at 7.63 GHz, respectively. To further investigate whether the source is resolved or not, we used the task \texttt{Modelfit} in DIFMAP to fit the radio emission component, but found no additional emission components in the residual map ($>$$3\sigma$). Therefore, \src remains compact and unresolved at the resolution of VLBA observations, with an upper limit on its size of $<$ 0.126 pc by using the deconvolved beam size of 7.63 GHz.

The brightness temperature of compact radio emission can be estimated as \citep[e.g.,][]{Ulvestad..2005}

\begin{equation}
T_b = 1.8 \times 10^9 \left( 1+z \right) \left( \frac{S_{\nu}}{1\, \text{mJy}} \right) \left( \frac{\nu}{1\, \text{GHz}} \right)^{-2} \left( \frac{\theta_1 \theta_2}{1\, \text{mas}^2} \right)^{-1}
\end{equation}

where S$_{\nu}$ is the peak flux density in mJy at the observing frequency $\nu$ in GHz, with $\theta_1$ and $\theta_2$ being the fitted Full Width at Half Maximum (FWHM) of the major and minor axes of the Gaussian component in units of milliarcseconds. Using the peak flux density and upper limit on the deconvolved source size derived from the VLBA 7.6 GHz image, we obtained a brightness temperature of $T_b\simgt$ 4.03 $\times$ 10$^{9}$ K. 
This $T_b$ limit significantly exceeds the brightness temperature threshold of normal star-formation process \citep[typically $\leq10^5$ K,][]{Condon1992}, suggesting that the VLBA component in AT2018cqh is clearly of non-thermal origin. 
The brightness temperature is
% however, comparable to 
lower than the equipartition temperature of $5 \times 10^{10}$ K \citep{Readhead1994}, 
% but much 
and also much less than the limiting brightness temperature of $\sim10^{12}$ K due to the inverse Compton catastrophe \citep{Kellermann..1969}.  
%which is usually used to indicate the intrinsic brightness temperature of the radio core in AGNs. 
This implies that the compact radio emission could not be associated with a jet, or the jet could have a low Doppler factor. Such a scenario could be further investigated with future VLBI observations to measure potential proper motion of the radio emission.

\subsection{Radio Light Curves: Continued Brightening }

Figure \ref{fig:radiolc} presents the radio light curves of AT2018cqh, at 0.89 GHz observed by ASKAP, 1.3-1.6 GHz observed by ASKAP, GMRT, e-MERLIN and VLBA, and 3 GHz observed by Very Long array (VLA).  
%Compared with \cite{Zhang..2024}, we have added new observational data including VLA 3 GHz results from three epochs between November 19, 2023 and September 4, 2024 (project code: 23B-321; PI: Collin Christy), ASKAP data since December 26, 2023, VLBA observations on August 6, 2024, and e-MERLIN data obtained on September 6, 2024. 
Note that in addition to data provided in Table 1 and \citet{Zhang..2024}, 
we retrieved the 3 GHz data observed by VLA (project code: 23B-321; PI: Collin Christy), consisting of three-epoch observations between Nov 19, 2023 and Sep 4, 2024. 
We then produced the calibrated clean images, and measured the flux densities 
at 3 GHz, following the procedures described in \citet{Yang..2022}. 
As shown in Figure \ref{fig:radiolc}, the 3 GHz flux displays a continuous rising trend. 
\cite{Zhang..2024} fitted the temporal evolution of the VLA Long Array Sky Survey (VLASS) 3 GHz flux with a power-law index $\alpha >$ 4.2, 
which becomes shallower ($\alpha$ = 2.23) if taking into account the data from the VLA follow-up observations at the same band. 
%In this work, by analyzing VLASS epoch III and three VLA epochs, we obtain a shallower temporal evolution with $\alpha$ = 2.23, indicating that the interaction between the outflow/jet and the circumnuclear medium (CNM) has weakened.
The light curve at 0.89 GHz observed by ASKAP exhibits a bump feature at $\delta$t$\approx$1846$-$2086 days relative to the time of the optical discovery. 
Following the bump in the light curve, the 0.89 GHz emission starts to rise again towards higher flux levels, 
but with fluctuations down to $\sim$2 months. 
To quantify the amplitude of the flux fluctuation, we perform a linear fit to the flux evolution starting at $\delta$t=2086 days, and then subtract this baseline to obtain variability amplitudes in the range of 7.4$\%$ - 27.5$\%$, 
with a median of 10.8$\%$. 
%Thanks to the higher cadence of 
%The ASKAP observations also reveal that 
%the most significant variability with an amplitude up to 109$\%$ and a characteristic timescale of 59 days. Such dramatic low-frequency variability could originate from ISS \citep{Goodwin..2023b}, intermittent ejection of the outflow/jet \citep{Romero-Canizales..2016, Perlman..2017, Mattila..2018}, or interaction between the outflow and multiple dense clouds \citep{Zhuang..2025}. We will discuss these possibilities in detail later.

It is well known that the effect of interstellar scintillation (ISS) could cause intraday radio variability in some AGNs with compact radio emission \citep{Lovell..2003, Rickett2007}, especially at lower frequencies. 
We investigate whether the observed radio variability in \src might be induced by the ISS effect, given the small source size of $\sim$134 micro-arcsecond ($\mu$as) at 7.6 GHz measured by VLBA (Section 3.1). 
Using the NE2001 electron density model \citep{Cordes&Lazio2022}, we infer that for the Galactic coordinates of AT2018cqh the transition frequency between strong and weak scintillation regimes occurs at $\sim$6.69 GHz and the angular size limit of the first Fresnel zone at the transition is 5.8 $\mu$as. 
%Using the \cite{Walker1998} formalism as appropriate for compact extragalactic sources, 
Adopting the formalisms provided in \cite{Walker1998} and the angular size of $\sim134$ $\mu$as obtained by VLBA , 
%it is expected that the radio emission from AT2018cqh could be in the strong, refractive scintillation regime if the source has an angular size of $\simlt$421$\mu$as. 
%Upper limit of Radii derived from VLBA of $3.9 \times 10^{17}$ cm at $D_A$ = 194 Mpc would correspond to angular diameters of 134$\mu$as ?? which band?. 
we estimate that the level of frequency-dependent random flux variations induced by the refractive ISS is from 32\% at 0.89\,GHz to 13\,\% at 3\,GHz for \srcs, with the time-scale for the modulation of  
%The emission from AT2018cqh is expected to be affected by ISS with a time-scale of variability of 
169 h (7 d) and 34 h (1.4 d), respectively. 
%a modulation fraction of 33\% at 0.89 GHz. 
%Although the cadence of ASKAP observations is insufficient to resolve shorter variability timescales, 
%This is comparable to
Given the comparable variability amplitudes, the flux fluctuations in the phase of continued radio brightening 
could be due to the ISS effect, though the cadence of ASKAP observations is insufficient to resolve the flux variability on shorter timescales. 
%The analysis of the ISS effect implies, however, that the bump in the 0.89 GHz light curve is 
The ISS effect, however, cannot be the primary factor causing the bump feature in the 0.89 GHz light curve, 
as its variability amplitude is $>$100\% and appears more regular. 

%the observed flux variability at 0.89 GHz far exceeds what would be expected from ISS effects. Therefore, we conclude that ISS is unlikely to be the primary cause of the significant flux variability at 0.89 GHz.

\subsection{Radio SED Evolution and Modeling }

\begin{figure}[t!]
\centering  
\includegraphics[width=0.46\textwidth]{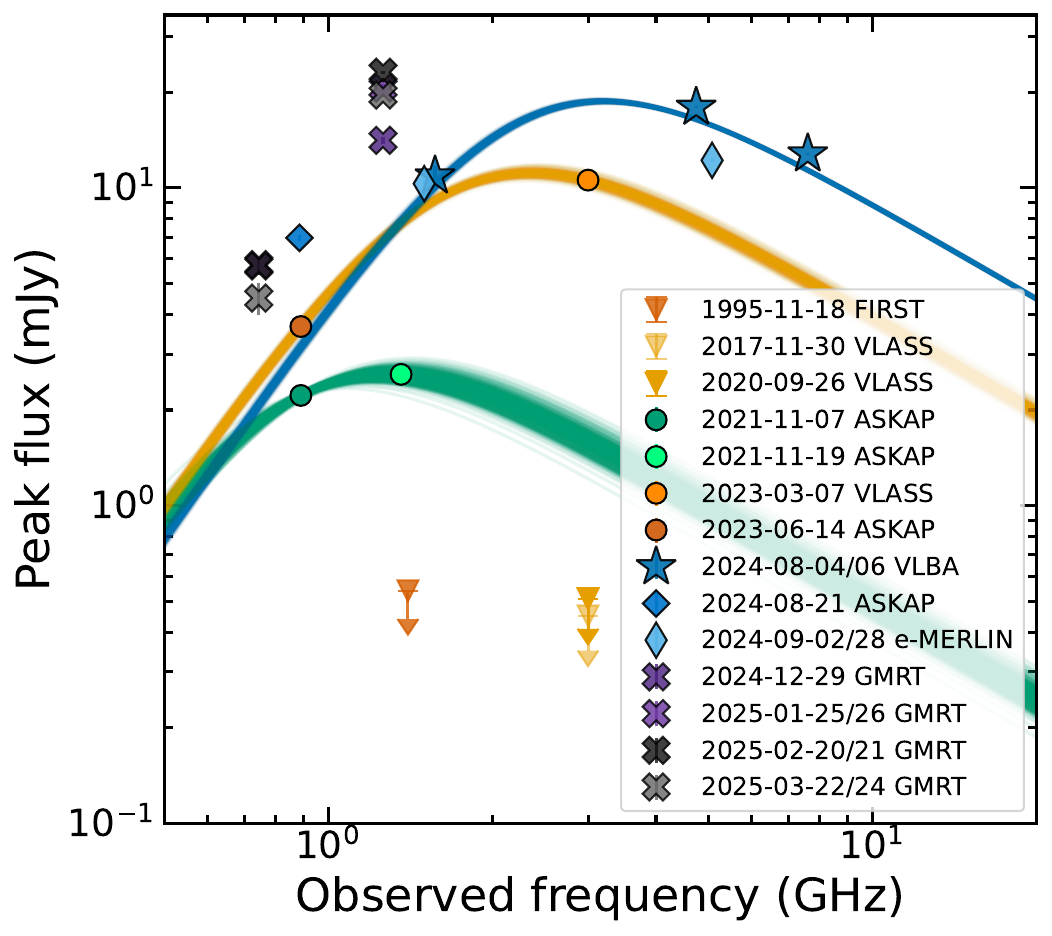}
\caption{The radio SEDs for three epochs that have quasi-simultaneous observations at different frequencies. For the nondetections, the corresponding 3$\sigma$ upper limits on flux density are shown. The green, orange and blue lines represent the best fit to each SED from our MCMC modeling (Section 3.3), which are the model realizations on a basis of 500 random samples from the MCMC chains. There is a steady rise in both peak flux density and frequency between the three epochs.
\label{fig:radiosed}}
\end{figure}

Figure \ref{fig:radiosed} displays the radio SEDs of \srcs. In addition to the data presented in \cite{Zhang..2024}, we have incorporated new measurements on the SEDs from quasi-simultaneous VLBA and e-MERLIN observations. 
Interestingly, although with higher spatial resolution, it is clear that the SED 
%shows a systematic shift 
has evolved to peak at higher frequency and flux density. %ies and peak frequencies.
%As in \cite{Goodwin..2022}, we 
To quantify the SED evolution, we fit the SED with a synchrotron emission
model in the standard framework of a {non-relativistic} outflow expanding
into and shocking the CNM with a density profile $n\propto r^{-k}$.  %outflow–CNM interaction, 
%following the same approach outlined in \cite{Goodwin..2022} and \cite{Zhang..2024}. 
%The interaction of outflow with the surrounding medium 
%leads to synchrotron emission owing to the acceleration of electrons 
%and amplification of magnetic fields.  
We use the synchrotron spectrum 2 described by \cite{Granot2002}, assuming $\nu_m \ll \nu_a$, where $\nu_m$ is the characteristic synchrotron frequency of the emitting electrons with the least energy and $\nu_a$ is the self-absorption frequency. 
Such a model has been widely used to explain the radio emission from non-relativistic TDEs \citep[e.g.,][]{Alexander..2016, Goodwin..2022}.

We then employ an Markov Chain Monte Carlo (MCMC) fitting technique \citep[python module \texttt{emcee}; ][]{Foreman-Mackey..2013} to marginalize over the synchrotron model parameters to determine the best-fitting parameters and uncertainties. %Due to the limited data points, 
Owing to the sparse sampling of the radio SED, especially at higher frequencies, we fix the synchrotron energy index in the optically thin regime to $p = 3$ \citep[e.g., ][]{Alexander..2016, Cendes..2021b}.
% In fact, we find that the derived parameters do not deviate significantly within the uncertainties if adopting other reasonable values, such as p ≈ 2–3.
In Figure \ref{fig:radiosed}, we show the resulting SED models, which provide reasonable fits to the data. From the best-fitting SED models, we determine the peak flux density and frequency, $F_{\nu, p}$ and $\nu_p$, respectively. 
We find that both $F_{\nu, p}$ and $\nu_p$ increase steadily with time, from 2.47 mJy and 1.42 GHz to 18.73 mJy and 3.22 GHz over a period of 997 days. 
Using the inferred values of $F_{\nu, p}$ and $\nu_p$, we can further adopt an equipartition analysis to derive the radius of the radio-emitting region ($R_{eq}$) and kinetic energy ($E_{eq}$) using the scaling relations outlined in \cite{BarniolDuran..2013}.
% \begin{equation}
% \begin{split}
% R_{eq} &= 1 \times 10^{17} \left( 21.8 \left( 525^{(p-1)} \right) \right)^{\frac{1}{13+2p}} \\
% &\quad \times \chi_e^{\frac{2-p}{13+2p}} F_{peak}^{\frac{6+p}{13+2p}} \left( \frac{d}{10^{28}\text{cm}} \right)^{\frac{2(p+6)}{13+2p}} \\ 
% &\quad \times \left( \frac{\nu_{peak}}{10 \text{GHz}} \right)^{-1} (1 + z)^{\frac{-19+3p}{13+2p}} f_A^{\frac{-5+p}{13+2p}} \\
% &\quad \times f_V^{\frac{-1}{13+2p}} 4^{\frac{1}{13+2p}} \xi^{\frac{1}{13+2p}} \text{cm}
% \end{split}
% \end{equation}
% \begin{equation}
% \begin{split}
% E_{eq} &= 1.3 \times 10^{48} \left( 21.8 \right)^{\frac{-2(p+1)}{13+2p}} \left( 525^{(p-1)} \chi_e^{(2-p)} \right)^{\frac{11}{13+2p}} \\
% &\quad \times F_{peak, mJy}^{\frac{14+3p}{13+2p}} \left( \frac{d}{10^{28}\text{cm}} \right)^{\frac{2(3p+14)}{13+2p}} \left( \frac{\nu_{peak}}{10 \text{GHz}} \right)^{-1} \left( 1+z \right)^{\frac{-27+5p}{13+2p}}\\ 
% &\quad \times f_A^{\frac{-3(p+1)}{13+2p}} f_V^{\frac{2(p+1)}{13+2p}} 4^{\frac{11}{13+2p}} \xi^{\frac{11}{13+2p}} \text{erg}
% \end{split}
% \end{equation}
Following the procedures described in \cite{Goodwin..2022}, we provide constraints for two different geometries, a spherical outflow and a mildly collimated conical outflow with a half-opening angle of $\phi = 30^{\circ}$, in order to account for the possible geometric dependence of outflow evolution.

\begin{figure*}[ht!]
\centering  
\includegraphics[width=0.76\textwidth]{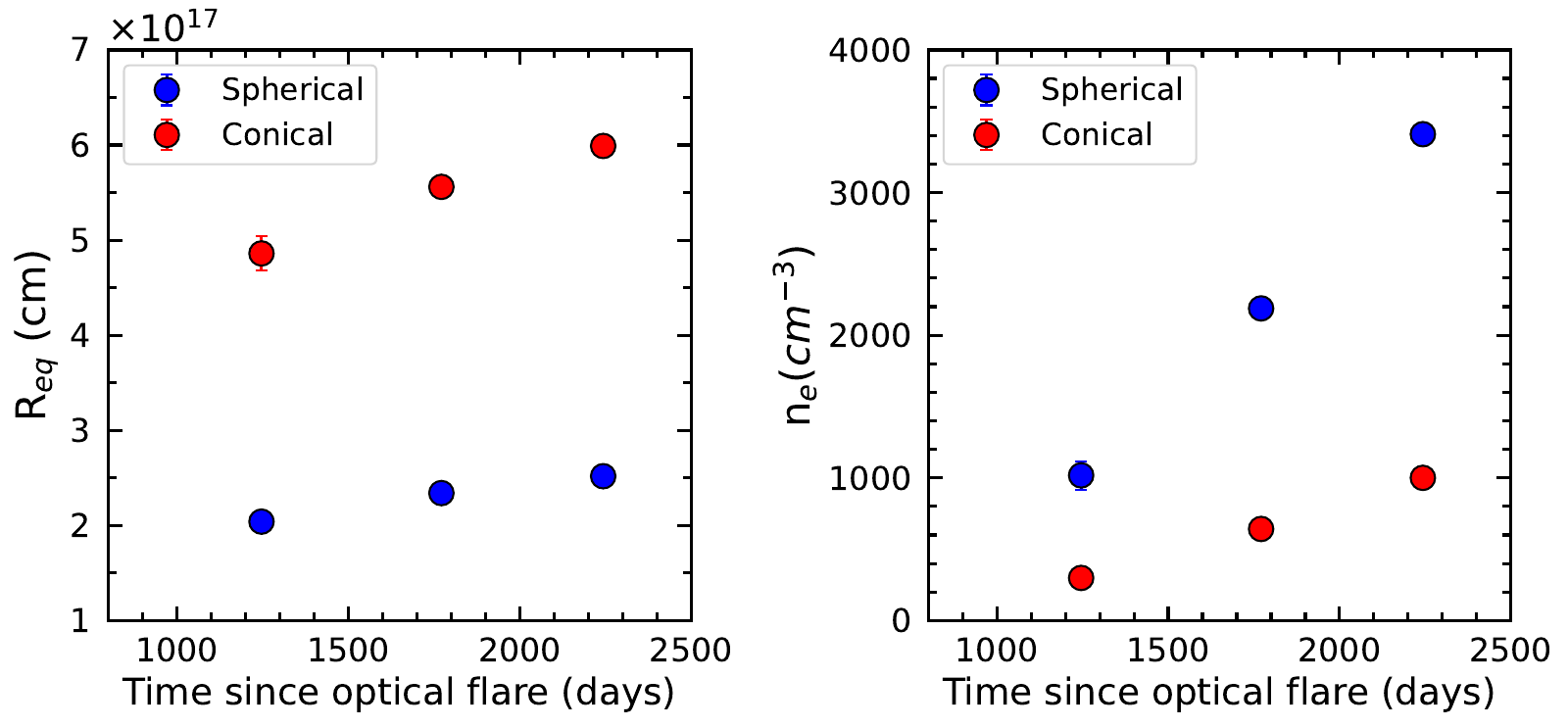}
\caption{The evolution of radius (left) and kinetic energy (right) as a function of time from our equipartition analysis, assuming a single outflow is launched into the CNM around the time of optical discovery. Blue and red filled circles indicate parameters for a spherical homogeneous and a collimated conical outflow, respectively.
\label{fig: parameter_plot}}
\end{figure*}

Through the equipartition analysis, we derive the temporal evolution of physical parameters in the radio-emitting region over three epochs.
% For the spherical outflow, we find that the radius are constact from $R_{eq} \approx 2.04 \times 10^{17}$ cm to $\approx 2.52 \times 10^{17}$ cm between t = 1246 and 2243 days, the energy and ambient density increases slightly from $E_{eq} \approx 1.0 \times 10^{50}$ erg to $\approx 5.62 \times 10^{50}$ erg and $n_e \approx 1017.6$ cm$^{-3}$ to $\approx 3409.1$ cm$^{-3}$ between t = 1246 and 2243 days, respectively.
As shown in Figure \ref{fig: parameter_plot}, for the spherical outflow, we find that the radius remains nearly constant, ranging from $R_{eq} \approx 2.04 \times 10^{17}$ cm to $\approx 2.52 \times 10^{17}$ cm between $\delta$t = 1246 and 2243 days, % Meanwhile, the energy and 
while the ambient density increases by a factor of $\sim$3, 
%$E_{eq}$ rising from $\approx 1.0 \times 10^{50}$ erg to $\approx 5.62 \times 10^{50}$ erg, and 
with $n_e$ increasing from $\approx 1017.6$ cm$^{-3}$ to $\approx 3409.1$ cm$^{-3}$ over the same period. 
Note that assuming a mildly collimated outflow yields a similar evolution in the parameters of $R_{eq}$ and $n_e$. 
%In the standard CNM shockwave model, 
Such an evolution in the equipartition radius and the ambient density is inconsistent with the continued radio brightening at low frequencies (Section 3.2), thus seems to disfavor the predictions of the standard CNM shockwave model. 

We note that the flux in $\sim$5 GHz observed by e-MERLIN on Sep 2, 2024 
is a factor of $\sim$1.5 lower than the 4.7 GHz flux observed by VLBA on Aug 6, 2024, 
while the flux in $\sim$1.5 GHz is consistent with each other. 
Such a flux difference cannot be due to the errors in flux measurements, as the target is bright. 
%This is ununusual if considering that  has a much higher spatial resolution ($>$?? times). 
The ISS-induced variability ($\sim5\%$) seems unlikely to explain the 5 GHz flux decrease observed by e-MERLIN.  
%as the expected ISS-induced variaiblity amplitude is $\sim$
Considering that the spatial resolution of the e-MERLIN observations 
is a factor of $\approx$30 larger than VLBA at the same band, the flux variability seems intrinsic to source and not due to the resolution effect. 
This suggests that there is a potentially slow evolution in the radio SED peaking towards lower frequencies on a time-scale of $\sim$1 month. 
On the other hand, the flux in the 0.89 GHz observed on Aug 21, 2024 exceeds that extrapolated from the best-fit SED model, by a factor of 2. 
Such an excess emission cannot be simply explained by the ISS effect, 
but represents intrinsic flux increase at the lower frequencies (see Figure \ref{fig:radiolc}).
More interestingly, the recent high-cadence GMRT observations at $\sim$0.7 and 1.3 GHz 
reveal a possibly new phase of fast flux rising. 
Further radio observations covering a broader frequency range are required to uncover whether there is a new pattern in the SED evolution, which is beyond the scope of current paper. 

%{\it Notably, the 0.89 GHz flux on 2024 Aug 21 shows a significant excess (by a factor of 2) compared to the latest SED. We therefore consider adding a second synchrotron component to fit the SED of this epoch (see Figure \ref{fig:twocomp}). The two-component synchrotron model provides a good fit to the latest SED, which suggests that there may be the presence of multiple synchrotron emission regions.
% , {\bf and the equipartition analysis yields physical parameters for both emission regions. (the detail parameters?)}
%Due to insufficient spectral sampling in earlier two epochs, we cannot determine whether their SEDs also require multiple synchrotron components. Both intermittent jet activity and interaction with multiple clouds remain plausible physical origins for the radio emission in AT2018cqh.}

\section{Discussions} 

In \citet{Zhang..2024}, we reported the detection of delayed radio emission from the TDE AT2018cqh, which might be related to the delayed launching of outflow. 
However, follow-up radio observing campaigns (including our e-MERLIN and VLBA observations) reveal an unusually evolving radio emission (up to $\delta$t$\sim$2510 days since its discovery). 
As mentioned in Section 3.2, the low-frequency radio emission displays continuous rising in flux with time.  
Particularly, there is a bump feature observed in the 0.89 GHz light curve between $\delta$t=1846 $-$ 2086 days, which cannot be explained under the standard model of an outflow interacting and shocking a homogeneous CNM. 
The rise to peak time of $\sim$115 days for the bump seems to disfavor a more complex model involving an inhomogeneous CNM, in which the time-scale is typically $\sim$days \citep{Matsumoto..2024}.  
%in which the radio emission originates from an i
The radio spectral evolution is also peculiar, characterized by 
%the radio SED of AT2018cqh has been shifting 
a shift toward higher peak frequency and peak flux density. 
This is in contrast to the radio evolution observed in other TDEs and in hydrodynamic simulations of the interaction between TDE outflow and a spherically symmetric CNM, where the peak frequency and flux density typically decrease over time \citep{Hu..2025}. 
While the increase in the peak flux density has been found in a few TDEs, 
the radio peak frequency does not vary or even decrease with time \citep{Cendes..2022, Cendes..2025, Christy..2024}, which is 
inconsistent with what we observed in \srcs. 
% whose peak frequency and peak flux density is decrease over time.
% \cite{Hu..2025} conducted hydrodynamic simulations of the radio emission produced by the interaction between a super-Eddington outflow, which generated from the disruption of a 1 M$_\odot$ main sequence star by a 10$^6$ M$_\odot$ black hole, and a spherically symmetric CNM. Their research revealed that the synthetic radio SED exhibited a peak frequency that shifted toward lower frequencies over time and consistent with the behavior observed in other radio-detected TDEs. However, this evolutionary trend markedly differs from the SED evolution observed in AT2018cqh.
%For AT2018hyz, the peak frequency of its SED evolved from 1.5GHz to 3.0GHz by 1200 days after optical detection, while the peak flux density remained nearly unchanged between the two epochs. \cite{Cendes..2022} suggested that the change in the SED properties during the latest epoch could be due to a combination of two outflows. 
%Additionally, the 0.89GHz lightcurve of AT2018cqh still exhibits fluctuations, a behavior rarely seen in other radio-detected TDEs. AT2023vwl showed a 25 per cent variation in its 1.5GHz flux density over just 9 days. \cite{Goodwin..2023a} proposed that this might be due to ISS effects rather than intrinsic variability.

%\textcolor{blue}
%{MOU suggests in blue: 
{On the other hand, based on the analysis using the equipartition theory (Section 3.3), we found that the equipartition radius increased very little over a period of $\sim$1,000 days between $\delta t=$1246 and 2243 days. This suggests that if the equipartition radius corresponds to the forward shock radius driven by the outflow in the CNM, its velocity during this period would be only about 0.018c for the spherical outflow case. 
By contrast, the shock velocity during the first 1246 days can be as high as 0.063c. 
If we take into account the possibility of a delayed outflow, the shock velocity would be even higher.
Based on the inferred CNM densities (1017.6 and 3409.1 cm$^{-3}$, Figure \ref{fig: parameter_plot}), we estimate that the kinetic energy of post shock CNM in the period of 1246--2243 days is about half of that in the first 1246 days. According to the Rankine-Hugoniot conditions, the internal energy of post-shock CNM is comparable to its kinetic energy. This indicates that about half of the energy is dissipated within the 1000 days. However, radiative cooling cannot dissipate this energy, as the cooling timescale \citep[exceeding 1000 years,][]{Sutherland1993} far exceeds the observed period. This challenges the self-consistency of the standard outflow-CNM model widely used in literature.}

\subsection{Outflow-cloud interaction}

As mentioned above, a notable feature of the radio evolution from AT2018cqh is the presence of fluctuations in the flux, 
particularly the bump feature. 
%\textcolor{red}
The bump feature indicates a significant drop in flux by a factor of 2 over a timescale of 2 months ($\delta t=2027 - 2086$ days), with the magnitude exceeding that caused by interstellar scintillation. This provides crucial clues for uncovering the physical origin of the radio flares. 
If it is generated by the outflow--CNM scenario, for affecting the overall radiation zone sized of $2\times 10^{17}$ cm (Figure \ref{fig: parameter_plot}), perturbations should go with a velocity higher than the light speed, suggesting that the conventional outflow--CNM scenario may lead to an inconsistent result.  
%While the bump-like flux fluctuation does not reconcile with the standard CNM shockwave model, 
%Considering the fact that the host of \src can be classified as type 2 
The aforementioned discrepancies could be naturally resolved by introducing the outflow--cloud interaction model: (1) outflow impacting multiple clouds \citep{Mou&Wang2021, Mou..2022, Zhuang..2025}; or (2) non-smooth outflow (with varying density) impacting one single cloud. 
%Either case would be expected 
The dense clouds are likely present in the circumnuclear environment of \srcs, as its host 
can be classified as a type 2 AGN \citep{Zhang..2024}. 
Thus, we examine the scenario of the outflow--cloud interaction by using hydrodynamic  simulations. 
%Here we explored two possible processes for generating the radio flares. One is the widely adopted outflow--CNM interaction scenario \citep{alexander2020}, and the other is outflow--cloud interaction \citep{mou2021b, mou2022}. 
%It's worth noting that, d
Due to the poor knowledge of the TDE outflow and the circumnuclear environment, we make simplifications to the physical parameters in simulations, while the model parameters that successfully fit the observations are not unique. 

\subsubsection{Model Settings} 
We conducted simulations with the ZEUS-3D code (\citealt{Clarke2010}), which includes the cosmic ray electrons (CRe) as a second fluid. For simplicity, we did not incorporate the magnetic field and diffusion of CRe. The hydrodynamic equations are
\begin{gather} 
\frac{\partial \rho}{\partial t} + \nabla \cdot (\rho {\bf v})=0 , \\
\rho \frac{d {\bf v}}{d t} = -\nabla (p_{1}+p_{2}) -\rho \nabla \Phi , \\
 \frac{\partial e_{1}}{\partial t} +\nabla \cdot(e_{1}{\bf v})=-p_{1}\nabla \cdot {\bf v}, \label{hydro3} \\
  \frac{\partial e_{2}}{\partial t}+\nabla \cdot(e_{2}{\bf v}) =-p_{2}\nabla \cdot {\bf v}, 
\end{gather}
where $p_1\equiv (\gamma_1-1)e_1$ is the thermal pressure ($\gamma_1=5/3$), $p_2\equiv (\gamma_2-1)e_2$ is the pressure of CRe ($\gamma_2=4/3$), and $\Phi$ is the gravitation potential ($\Phi=-GM_{\rm bh}/r$).
We did not simulate the acceleration process of electrons, but instead, directly injected the CRe component as the second fluid which is embodied as the energy density $e_2$. As mentioned in the next subsection, $e_2$ is assigned the value of $\frac{5}{3}$ $\epsilon_e$$e_{d}$, where $e_{d}$ is thermal pressure in the downstream. The shock front in ZEUS-3D is typically captured with 4 meshes, and thus we inject the CRe at the 4th-mesh when the parameters of the post-shock gas have been stabilized. 
The simulation domain is 2.5 dimensional spherical coordinates, of which the system is symmetric in $\phi-$direction. 
The initial density distribution of the hot CNM is assumed in a form of $\rho_{\rm cnm}(r)=\rho_0 r^{-n} $. 
Here we set it to be the same as the density distribution of the Galactic Center, i.e., $\rho_{\rm cnm}(r)=3 ~{\rm m_{H} ~ cm^{-3}} (r/{\rm pc})^{-1}$ \citep{Gillessen..2019}. 

Using the radio SED from three epochs, and applying the minimal energy method \citep{BarniolDuran..2013}, we find that the sizes of the radiation zones of three epochs ($\delta t=$1246, 1730, and 2243 days in the observer's frame) show tiny difference despite a time span of 1000 days. Thus, it is possible that the outflow continuously interacts with one single cloud. By performing a serial of simulations, we find that such a scenario successfully reproduces the radio flares of AT2018cqh. Specifically, the cloud is assumed to be spherical shape with a radius of 0.08 pc, and its center is placed at the polar axis with a distance of 0.43 pc from the SMBH. The TDE outflow is injected at the inner boundary of the simulation domain for $t\leq 1.6$ years, and is confined within two cones along the bipolar directions, whose total solid angle is $0.1\times 4\pi$ sr (corresponding to a half-opening angle of 25.8$^{\circ}$). The outflow velocity decreases linearly from 0.5 c to 0.2 c, and the mass outflow rate increases with time:
\begin{equation}
\dot{M}_{\rm out}(t) = 0.006 M_{\odot} {\rm yr^{-1}} \cdot \left[ 2 + 100 \times \left( \frac{t}{1.6\, {\rm yr}} \right)^3 \right] .
\end{equation}

\begin{figure*}[tph!] 
\centering
\includegraphics[width=\linewidth]{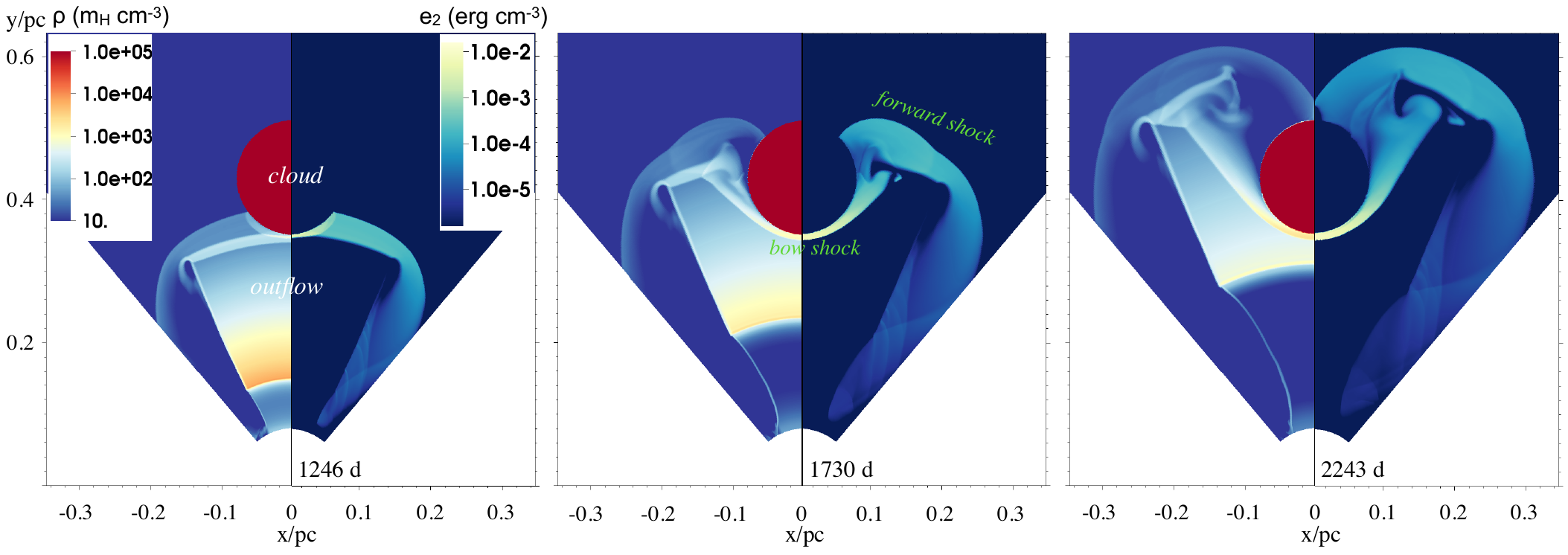}%{simufig1.pdf} 
\caption{Evolution for the outflow--cloud interaction. Panels from the \emph{left} to the \emph{right} show the density and CRe energy at three different epochs (observer's frame): $\delta t=1246$ d, 1730 d and 2243 d.}
\label{simufig1}
\end{figure*}

\subsubsection{Relativistic Electrons and Magnetic Field}
The electron acceleration efficiency $\epsilon_e$ is assumed to be the fraction of the energy flux that can be dissipated at the shock surface (i.e., the change in the kinetic energy flux across the shock) channelled into the accelerated relativistic electrons in the downstream. 
In the frame of shock front, the acceleration efficiency is  % $\epe$ is 
\begin{equation}
\epsilon_e=\frac{e_2 v_{d}}{\frac{1}{2}\rho_{i} v^3_s (1-C^{-2})}
\end{equation}
where $e_2$ is the energy density of CRe in the downstream, $v_{d}$ is the downstream velocity, $\rho_{i}$ is the pre-shock density, and $C=4\mathcal{M}^2/(\mathcal{M}^2+3)$ is the compression ratio. When the Mach number $\mathcal{M} \gg 1$, we have $v_{d}=v_s/4$ and $C=4$. In this case, the above equation can be further simplified as $\epsilon_e=0.6 e_2/e_{d}$, where $e_{d}$ is thermal pressure in the downstream.  

We did not include the magnetic field in the simulations, and simply assumed that the energy ratio of magnetic field to CRe in each mesh remains constant, i.e., $[B^2/(8\pi)]/e_2=\epsilon_b/\epsilon_e=$constant. We set $\epsilon_b=\epsilon_e=0.1$ in this study. 
We present the density distribution and CRe's energy density distribution for different moments.  

\begin{figure}[h!]
\includegraphics[width=0.9\columnwidth]{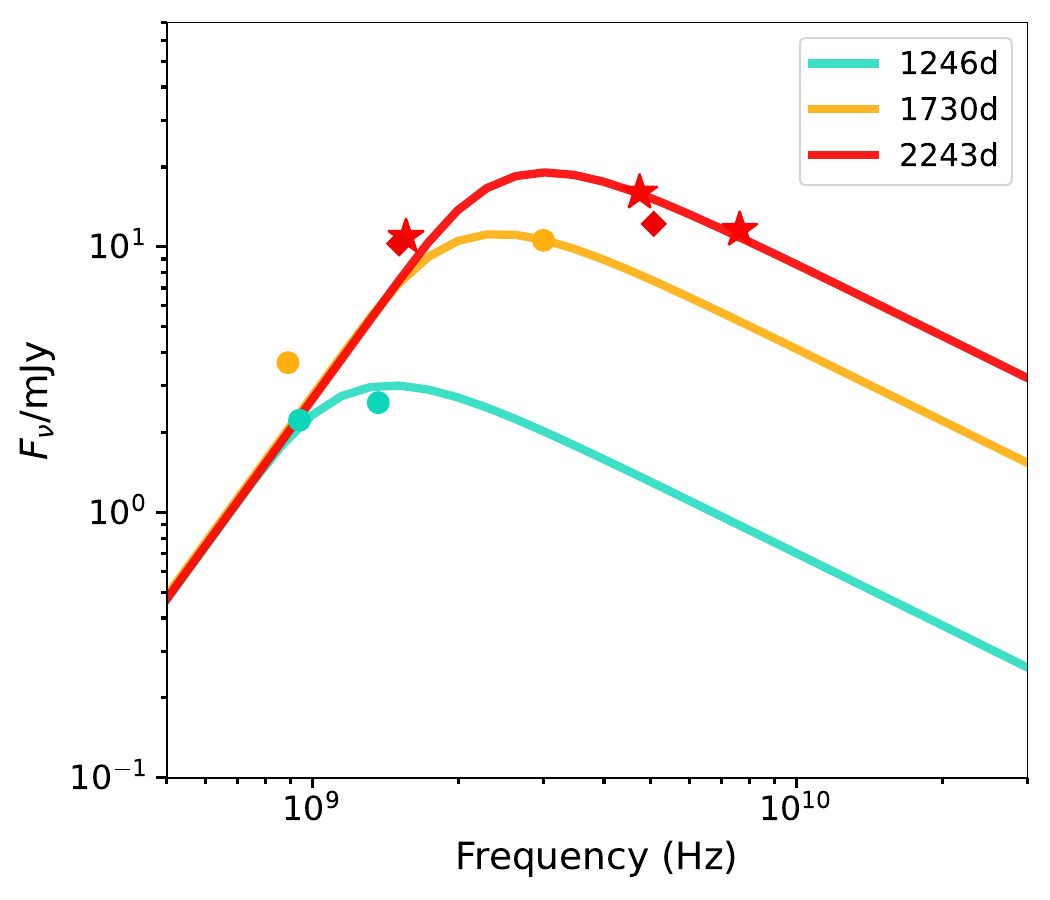}%{Fig7j0233sed.pdf}
 \caption{Hydrodynamic modeling of the radio SEDs of \srcs. The cyan, orange and red lines correspond to the model SEDs at $\delta t=1246$ d, 1730 d and 2243 d in observer's frame, respectively. The filled circles, stars and diamonds represent the observed flux densities, as shown in Figure \ref{fig:radiosed}.} 
 \label{simufig2}
\end{figure}

%\begin{figure}[h!]
%\includegraphics[width=0.9\columnwidth]{Fig8F09GHz.pdf}
% \caption{The simulated variation of the 0.9 GHz radio flux in arbitrary unit. In this test, we applied random perturbations only to the mass flux at the inner boundary, while keeping all other parameters unchanged.}
% \label{simufig3}
%\end{figure}

\subsubsection{Simulation Results}
Our simulations can provide the distribution of CR and the magnetic field strength at any given moment, thereby yielding the radio spectrum at that moment. However, the radiation zone formed by the bow shock is not spherically symmetric, and the radio SED varies with line of sight. Here we simplify the situation by treating the radiation zone as a radiation shell surrounding the spherical cloud, where the CRe's energy density and magnetic field strength inside are uniform. According to hydrodynamic simulations (see Figure. \ref{simufig1}), the radiation shell is thin compared to the cloud, with an approximate thickness of $\sim 0.01$ pc. We assume, approximately, that the radiative zone is a thin spherical shell with a radius of 0.09 pc and a thickness of 0.01 pc.  
The CRes per unit volume follow a power-law distribution: $dn(E)/dE=A E^{-p}$, where the coefficient $A \propto e_2$, and we fix the index $p=2.8$. The synchrotron emission spectrum per unit volume can be computed as: 
\begin{equation}
j_{\nu} \propto A B^{\frac{p+1}{2}} 
\propto \left(8\pi \frac{\epsilon_b}{\epsilon_e} \right)^{\frac{p+1}{4}} e^{\frac{p+5}{4}}_2  
\end{equation}
To obtain the total synchrotron radiation from the volume integration, we use the following approximation: $\int \left(8\pi \frac{\epsilon_b}{\epsilon_e} \right)^{\frac{p+1}{4}} e^{\frac{p+5}{4}}_2 dV=\bar{B}^{\frac{p+1}{2}} \int e_2 dV $,
where the equivalent magnetic field strength is
\begin{equation}
\bar{B}=\left[ \left(8\pi \frac{\epsilon_b}{\epsilon_e} \right)^{\frac{p+1}{4}} \cdot \frac{\int e^{\frac{p+5}{4}}_2 dV}{\int e_2 dV} \right]^{\frac{2}{p+1}}
\end{equation}
\smallskip

Using such a simplified radiation shell, the total energy of CRe from simulations ($\int e_2 dV$) and the equivalent magnetic field strength $\bar{B}$, we calculate the self-absorbed radio spectrum. The radio SEDs from our simulations are shown in Figure \ref{simufig2}, which can match the observed radio SEDs at three epochs for the outflow-cloud-interaction picture. 
%and we find that the outflow-cloud scenario can match the radio SED at three epochs. 

%\textcolor{blue}{MOU: In addition, we also investigated the potential intrinsic causes of the radio flux variability. Building on the simulations mentioned above, we introduced random perturbations to the mass flux at the inner boundary by multiplying it (Equation 6) with an oscillation factor $f_M(t)$. This factor changes every 20 days, taking a random value between 0 and 2. Figure \ref{simufig3} presents one of the simulation tests, and we find that the radio flux at 0.9 GHz can vary within 3 months for our simulation setup. }

%\textcolor{blue}{
Through simulations, we demonstrate that the SED evolution of AT2018cqh can be constrained within the theoretical framework of outflow-cloud interaction. 
The size of the emission region -- that is, the size of the bow shock — is roughly comparable to the size of the cloud, and therefore shows no significant change over the duration of 1000 days (from 1246 to 2243 days). 
This is also consistent with the absence of a resolved component in the VLBA imaging observations. %Additionally, the rapid variability of AT2018cqh on monthly timescale may be related to fluctuations in the mass outflow rate. }
%This result indicate that, the delayed radio flares in J0233 does not necessarily 
%We further examined the possibility that the density disturbances caused by the outflow itself lead to fluctuations in the radio flux. 
%At the inner boundary, we applied a sine function perturbation to the density of the TDE outflow, with a period of 60 days.
% {\bf outflow-cloud model can give the velocity of outflow? (The velocity derived from VLBA is 0.13c)}
% {\bf outflow-cloud model can fitting the flux exceed of 0.89 GHz in latest SED and the oscillation of light-curve?}

\section{Conclusions}
We present follow-up VLBA and e-MERLIN radio observations of AT2018cqh extending to $\delta$t $\sim$ 2250 days post discovery. %, as well as analyzed the ongoing ASKAP archive data. 
With the high resolution imaging at a mas-scale, VLBA reveals a compact radio emission unresolved at a physical scale of $\simlt$0.13 pc at 7.6 GHz, with a brightness temperature of $T_b$ $\simgt$ 4 $\times$ 10$^{9}$ K. In addition to the core, e-MERLIN observation at 5 GHz does not detect any extended emission components at a scale of 34.6 pc, supporting the compactness of the radio emission. 
The quasi-simultaneous VLBA and e-MERLIN observations allow for constraining the radio spectral evolution at later times, and find a shift in the SED 
%Meanwhile, the SED continue to shift 
to a higher peak flux density and frequency. 
Such a radio SED evolution is unprecedented among known TDEs. 
The radio light curve at 0.89 GHz observed with ASKAP suggests a continued flux increase, with a bump feature at $\delta t \approx$1846 days. 
The radio flux and SED evolution properties of \src are inconsistent with the predictions of conventional outflow-CNM interaction model, but can be naturally explained by the outflow-cloud interaction.  
%outflow-cloud interaction. 
We verified the latter model by hydrodynamic simulations and found that it can reproduce the peculiar radio SED evolution observed in \srcs. 

%\textbf{
%Moreover, the interaction between outflow and clouds in AT2018cqh can be verified through the X-ray light curve and spectral properties\citep{Mou..2021}.}
Since the most recent GMRT observations at $\sim$0.7 and 1.3 GHz 
find a possibly new phase of fast flux rising, 
continued radio observations covering a broader frequency range are required to 
catch the peak of the radio SED, uncovering whether there is a new pattern in the SED evolution.  
%Due to its peculiar evolution of the delayed radio flare, \src might represent the best case known so far to test the 
%We propose that the synchronous increase in the peak flux density with peak frequency in the radio spectral evolution could be a signpost of outflow-cloud interaction, which can be tested with future radio observations of TDEs, especially those occurred in AGNs. 
%Future observations of additional TDEs with similar radio evolution properties 
% which can be used to test in detail the outflow-cloud interaction model by comparing with 
% hydrodynamic simulations. 
With new radio observations, the outflow–cloud interaction model can be tested in detail through comparisons with hydrodynamic simulations.
In addition to radio flares, theoretical studies demonstrate that the outflow–cloud interaction is capable of generating X-ray emission when the cloud shock propagates through the cloud \citep{Mou..2021}. Based on our simulation data (Section 4.1.1), the intrinsic X-ray luminosity in 0.3--10 keV range is inferred to be $L_{\rm X}\approx10^{41}$~\erg, which is an order of magnitude lower than those obtained from the recent XMM-Newton and Swift/XRT observations (Wang et al. 2025, private communication). The X-ray spectrum of \src is currently dominated by a thermal blackbody component ($\rm kT\approx60~eV$) likely powered by the accretion process. 
The outflow-cloud interaction model could be further tested with future sensitive X-ray observations, once the TDE's thermal disk emission has decayed. %, leaving only an X-ray spectrum dominated by a power-law component. 

\acknowledgments{
We thank the anonymous referee for the constructive suggestions and detailed comments, which helped to improve the clarity and quality of this paper.
We thank the staff of the e-MERLIN, VLBA and GMRT, that made these observations possible. 
e-MERLIN is a National Facility operated by the University of Manchester at Jodrell Bank Observatory on behalf of STFC.
The National Radio Astronomy Observatory is a facility of the National Science Foundation operated under cooperative agreement by Associated Universities, Inc. 
GMRT is run by the National Centre for Radio Astrophysics of the Tata Institute of Fundamental Research. 
%The National Radio Astronomy Observatory is a facility of the National Science Foundation operated under cooperative agreement by Associated Universities, Inc.
%The European VLBI Network is a joint facility of independent European, African, Asian, and North American radio astronomy institutes. Scientific results from data presented in this publication are derived from the following EVN project code: ES091. 
The Australian SKA Pathfinder is
part of the Australia Telescope National Facility, which is
managed by CSIRO. Operation of ASKAP is funded by the
Australian Government with support from the National
Collaborative Research Infrastructure Strategy. This paper
includes archived data obtained through the CSIRO ASKAP
Science Data Archive, CASDA (http://data.csiro.au). 
%This research makes use of data products from the Wide-field Infrared Survey Explorer, which is a joint project of the University of California, Los Angeles, and the Jet Propulsion Laboratory/California Institute of Technology, funded by the National Aeronautics and Space Administration. 
The work is supported by the National SKA program of China (2022SKA0130102) 
and the National Science Foundation of China (NSFC) through grant No. 12192220, 12192221.  
%and National SKA program of China (2022SKA0130102). 
%the National SKA Program of China (2022SKA0130102)
%SKA Fast Radio Burst and High-Energy Transients Project (2022SKA0130102). 
%, and the National Science Foundation of China (NSFC) through grant No. 12192220, 12192221, 11988101. 
X.S. acknowledges the science research grants from the China Manned Space Project with NO. CMS-CSST-2025-A07. 
G.M. was supported by the NSFC (No. 12473013, 12133007).
{Y.X. was supported by the NSFC (No. 12025303).}
}
\smallskip
\smallskip

 \bibliographystyle{aasjournal}
\bibliography{AT2018cqh-2_radio.bib}

\begin{thebibliography}{}
\expandafter\ifx\csname natexlab\endcsname\relax\def\natexlab#1{#1}\fi
\providecommand{\url}[1]{\href{#1}{#1}}
\providecommand{\dodoi}[1]{doi:~\href{http://doi.org/#1}{\nolinkurl{#1}}}
\providecommand{\doeprint}[1]{\href{http://ascl.net/#1}{\nolinkurl{http://ascl.net/#1}}}
\providecommand{\doarXiv}[1]{\href{https://arxiv.org/abs/#1}{\nolinkurl{https://arxiv.org/abs/#1}}}

\bibitem[{{Alexander} {et~al.}(2016){Alexander}, {Berger}, {Guillochon},
  {Zauderer}, \& {Williams}}]{Alexander..2016}
{Alexander}, K.~D., {Berger}, E., {Guillochon}, J., {Zauderer}, B.~A., \&
  {Williams}, P.~K.~G. 2016, \apjl, 819, L25,
  \dodoi{10.3847/2041-8205/819/2/L25}

\bibitem[{{Alexander} {et~al.}(2020){Alexander}, {van Velzen}, {Horesh}, \&
  {Zauderer}}]{Alexander..2020}
{Alexander}, K.~D., {van Velzen}, S., {Horesh}, A., \& {Zauderer}, B.~A. 2020,
  \ssr, 216, 81, \dodoi{10.1007/s11214-020-00702-w}

\bibitem[{{Alexander} {et~al.}(2017){Alexander}, {Wieringa}, {Berger},
  {Saxton}, \& {Komossa}}]{Alexander..2017}
{Alexander}, K.~D., {Wieringa}, M.~H., {Berger}, E., {Saxton}, R.~D., \&
  {Komossa}, S. 2017, \apj, 837, 153, \dodoi{10.3847/1538-4357/aa6192}

\bibitem[{{Andreoni} {et~al.}(2022){Andreoni}, {Coughlin}, {Perley}, {Yao},
  {Lu}, {Cenko}, {Kumar}, {Anand}, {Ho}, {Kasliwal}, {de Ugarte Postigo},
  {Sagu{\'e}s-Carracedo}, {Schulze}, {Kann}, {Kulkarni}, {Sollerman}, {Tanvir},
  {Rest}, {Izzo}, {Somalwar}, {Kaplan}, {Ahumada}, {Anupama}, {Auchettl},
  {Barway}, {Bellm}, {Bhalerao}, {Bloom}, {Bremer}, {Bulla}, {Burns},
  {Campana}, {Chandra}, {Charalampopoulos}, {Cooke}, {D'Elia}, {Das}, {Dobie},
  {Ag{\"u}{\'\i} Fern{\'a}ndez}, {Freeburn}, {Fremling}, {Gezari}, {Goode},
  {Graham}, {Hammerstein}, {Karambelkar}, {Kilpatrick}, {Kool}, {Krips},
  {Laher}, {Leloudas}, {Levan}, {Lundquist}, {Mahabal}, {Medford}, {Miller},
  {M{\"o}ller}, {Mooley}, {Nayana}, {Nir}, {Pang}, {Paraskeva}, {Perley},
  {Petitpas}, {Pursiainen}, {Ravi}, {Ridden-Harper}, {Riddle}, {Rigault},
  {Rodriguez}, {Rusholme}, {Sharma}, {Smith}, {Stein}, {Th{\"o}ne},
  {Tohuvavohu}, {Valdes}, {van Roestel}, {Vergani}, {Wang}, \&
  {Zhang}}]{Andreoni..2022}
{Andreoni}, I., {Coughlin}, M.~W., {Perley}, D.~A., {et~al.} 2022, \nat, 612,
  430, \dodoi{10.1038/s41586-022-05465-8}

\bibitem[{{Armijos-Abenda{\~n}o} {et~al.}(2022){Armijos-Abenda{\~n}o},
  {L{\'o}pez}, {Llerena}, \& {Logan}}]{Armijos2022}
{Armijos-Abenda{\~n}o}, J., {L{\'o}pez}, E., {Llerena}, M., \& {Logan},
  C.~H.~A. 2022, \mnras, 514, 1535, \dodoi{10.1093/mnras/stac1442}

\bibitem[{{Barniol Duran} {et~al.}(2013){Barniol Duran}, {Nakar}, \&
  {Piran}}]{BarniolDuran..2013}
{Barniol Duran}, R., {Nakar}, E., \& {Piran}, T. 2013, \apj, 772, 78,
  \dodoi{10.1088/0004-637X/772/1/78}

\bibitem[{{Blanchard} {et~al.}(2017){Blanchard}, {Nicholl}, {Berger},
  {Guillochon}, {Margutti}, {Chornock}, {Alexander}, {Leja}, \&
  {Drout}}]{Blanchard..2017}
{Blanchard}, P.~K., {Nicholl}, M., {Berger}, E., {et~al.} 2017, \apj, 843, 106,
  \dodoi{10.3847/1538-4357/aa77f7}

\bibitem[{{Brown} {et~al.}(2015){Brown}, {Levan}, {Stanway}, {Tanvir}, {Cenko},
  {Berger}, {Chornock}, \& {Cucchiaria}}]{Brown..2015}
{Brown}, G.~C., {Levan}, A.~J., {Stanway}, E.~R., {et~al.} 2015, \mnras, 452,
  4297, \dodoi{10.1093/mnras/stv1520}

\bibitem[{{Bu} {et~al.}(2023){Bu}, {Chen}, {Mou}, {Qiao}, \&
  {Yang}}]{Bu..2023a}
{Bu}, D.-F., {Chen}, L., {Mou}, G., {Qiao}, E., \& {Yang}, X.-H. 2023, \mnras,
  521, 4180, \dodoi{10.1093/mnras/stad804}

\bibitem[{{Burrows} {et~al.}(2011){Burrows}, {Kennea}, {Ghisellini}, {Mangano},
  {Zhang}, {Page}, {Eracleous}, {Romano}, {Sakamoto}, {Falcone}, {Osborne},
  {Campana}, {Beardmore}, {Breeveld}, {Chester}, {Corbet}, {Covino},
  {Cummings}, {D'Avanzo}, {D'Elia}, {Esposito}, {Evans}, {Fugazza}, {Gelbord},
  {Hiroi}, {Holland}, {Huang}, {Im}, {Israel}, {Jeon}, {Jeon}, {Jun}, {Kawai},
  {Kim}, {Krimm}, {Marshall}, {P. M{\'e}sz{\'a}ros}, {Negoro}, {Omodei},
  {Park}, {Perkins}, {Sugizaki}, {Sung}, {Tagliaferri}, {Troja}, {Ueda},
  {Urata}, {Usui}, {Antonelli}, {Barthelmy}, {Cusumano}, {Giommi}, {Melandri},
  {Perri}, {Racusin}, {Sbarufatti}, {Siegel}, \& {Gehrels}}]{Burrows..2011}
{Burrows}, D.~N., {Kennea}, J.~A., {Ghisellini}, G., {et~al.} 2011, \nat, 476,
  421, \dodoi{10.1038/nature10374}

\bibitem[{{Bykov} {et~al.}(2024){Bykov}, {Gilfanov}, \&
  {Sunyaev}}]{Bykov..2024}
{Bykov}, S.~D., {Gilfanov}, M.~R., \& {Sunyaev}, R.~A. 2024, \mnras, 527, 1962,
  \dodoi{10.1093/mnras/stad3355}

\bibitem[{{Cendes} {et~al.}(2021){Cendes}, {Alexander}, {Berger}, {Eftekhari},
  {Williams}, \& {Chornock}}]{Cendes..2021b}
{Cendes}, Y., {Alexander}, K.~D., {Berger}, E., {et~al.} 2021, \apj, 919, 127,
  \dodoi{10.3847/1538-4357/ac110a}

\bibitem[{{Cendes} {et~al.}(2022){Cendes}, {Berger}, {Alexander}, {Gomez},
  {Hajela}, {Chornock}, {Laskar}, {Margutti}, {Metzger}, {Bietenholz},
  {Brethauer}, \& {Wieringa}}]{Cendes..2022}
{Cendes}, Y., {Berger}, E., {Alexander}, K.~D., {et~al.} 2022, \apj, 938, 28,
  \dodoi{10.3847/1538-4357/ac88d0}

\bibitem[{{Cendes} {et~al.}(2024){Cendes}, {Berger}, {Alexander}, {Chornock},
  {Margutti}, {Metzger}, {Wieringa}, {Bietenholz}, {Hajela}, {Laskar}, {Stroh},
  \& {Terreran}}]{Cendes..2024}
---. 2024, \apj, 971, 185, \dodoi{10.3847/1538-4357/ad5541}

\bibitem[{{Cendes} {et~al.}(2025){Cendes}, {Berger}, {Beniamini}, {Gill},
  {Matsumoto}, {Alexander}, {Bietenholz}, {Hajela}, {Christy}, {Chornock},
  {Gomez}, {Gurwell}, {Keating}, {Laskar}, {Margutti}, {Rao}, {Velez}, \&
  {Wieringa}}]{Cendes..2025}
{Cendes}, Y., {Berger}, E., {Beniamini}, P., {et~al.} 2025, arXiv e-prints,
  arXiv:2507.08998, \dodoi{10.48550/arXiv.2507.08998}

\bibitem[{{Chan} {et~al.}(2019){Chan}, {Piran}, {Krolik}, \&
  {Saban}}]{Chan..2019}
{Chan}, C.-H., {Piran}, T., {Krolik}, J.~H., \& {Saban}, D. 2019, \apj, 881,
  113, \dodoi{10.3847/1538-4357/ab2b40}

\bibitem[{{Christy} {et~al.}(2024){Christy}, {Alexander}, {Margutti},
  {Wieringa}, {Cendes}, {Chornock}, {Laskar}, {Berger}, {Bietenholz},
  {Coppejans}, {De Colle}, {Eftekhari}, {Holoien}, {Matsumoto}, {Miller-Jones},
  {Ramirez-Ruiz}, {Saxton}, \& {van Velzen}}]{Christy..2024}
{Christy}, C.~T., {Alexander}, K.~D., {Margutti}, R., {et~al.} 2024, \apj, 974,
  18, \dodoi{10.3847/1538-4357/ad675b}

\bibitem[{{Clarke}(2010)}]{Clarke2010}
{Clarke}, D.~A. 2010, \apjs, 187, 119, \dodoi{10.1088/0067-0049/187/1/119}

\bibitem[{{Condon}(1992)}]{Condon1992}
{Condon}, J.~J. 1992, \araa, 30, 575,
  \dodoi{10.1146/annurev.aa.30.090192.003043}

\bibitem[{{Cordes} \& {Lazio}(2002)}]{Cordes&Lazio2022}
{Cordes}, J.~M., \& {Lazio}, T.~J.~W. 2002, arXiv e-prints, astro,
  \dodoi{10.48550/arXiv.astro-ph/0207156}

\bibitem[{{Cornwell} {et~al.}(2008){Cornwell}, {Golap}, \&
  {Bhatnagar}}]{Cornwell..2008}
{Cornwell}, T.~J., {Golap}, K., \& {Bhatnagar}, S. 2008, IEEE Journal of
  Selected Topics in Signal Processing, 2, 647,
  \dodoi{10.1109/JSTSP.2008.2005290}

\bibitem[{{De Colle} {et~al.}(2012){De Colle}, {Guillochon}, {Naiman}, \&
  {Ramirez-Ruiz}}]{DeColle..2012}
{De Colle}, F., {Guillochon}, J., {Naiman}, J., \& {Ramirez-Ruiz}, E. 2012,
  \apj, 760, 103, \dodoi{10.1088/0004-637X/760/2/103}

\bibitem[{{Deller} {et~al.}(2011){Deller}, {Brisken}, {Phillips}, {Morgan},
  {Alef}, {Cappallo}, {Middelberg}, {Romney}, {Rottmann}, {Tingay}, \&
  {Wayth}}]{Deller..2011}
{Deller}, A.~T., {Brisken}, W.~F., {Phillips}, C.~J., {et~al.} 2011, \pasp,
  123, 275, \dodoi{10.1086/658907}

\bibitem[{{Fangyi} {et~al.}(2025){Fangyi}, {Hu}, {Goodwin}, {Price}, {Mandel},
  {Sari}, \& {Hayasaki}}]{Hu..2025}
{Fangyi}, {Hu}, {Goodwin}, A., {et~al.} 2025, arXiv e-prints, arXiv:2507.01273,
  \dodoi{10.48550/arXiv.2507.01273}

\bibitem[{{Foreman-Mackey} {et~al.}(2013){Foreman-Mackey}, {Hogg}, {Lang}, \&
  {Goodman}}]{Foreman-Mackey..2013}
{Foreman-Mackey}, D., {Hogg}, D.~W., {Lang}, D., \& {Goodman}, J. 2013, \pasp,
  125, 306, \dodoi{10.1086/670067}

\bibitem[{{Gezari}(2021)}]{Gezari2021}
{Gezari}, S. 2021, \araa, 59, 21, \dodoi{10.1146/annurev-astro-111720-030029}

\bibitem[{{Gillessen} {et~al.}(2019){Gillessen}, {Plewa}, {Widmann}, {von
  Fellenberg}, {Schartmann}, {Habibi}, {Jimenez Rosales}, {Baub{\"o}ck},
  {Dexter}, {Gao}, {Waisberg}, {Eisenhauer}, {Pfuhl}, {Ott}, {Burkert}, {de
  Zeeuw}, \& {Genzel}}]{Gillessen..2019}
{Gillessen}, S., {Plewa}, P.~M., {Widmann}, F., {et~al.} 2019, \apj, 871, 126,
  \dodoi{10.3847/1538-4357/aaf4f8}

\bibitem[{{Goodwin} {et~al.}(2022){Goodwin}, {van Velzen}, {Miller-Jones},
  {Mummery}, {Bietenholz}, {Wederfoort}, {Hammerstein}, {Bonnerot}, {Hoffmann},
  \& {Yan}}]{Goodwin..2022}
{Goodwin}, A.~J., {van Velzen}, S., {Miller-Jones}, J.~C.~A., {et~al.} 2022,
  \mnras, 511, 5328, \dodoi{10.1093/mnras/stac333}

\bibitem[{{Goodwin} {et~al.}(2023{\natexlab{a}}){Goodwin}, {Miller-Jones}, {van
  Velzen}, {Bietenholz}, {Greenland}, {Cenko}, {Gezari}, {Horesh}, {Sivakoff},
  {Yan}, {Yu}, \& {Zhang}}]{Goodwin..2023a}
{Goodwin}, A.~J., {Miller-Jones}, J.~C.~A., {van Velzen}, S., {et~al.}
  2023{\natexlab{a}}, \mnras, 518, 847, \dodoi{10.1093/mnras/stac3127}

\bibitem[{{Goodwin} {et~al.}(2023{\natexlab{b}}){Goodwin}, {Alexander},
  {Miller-Jones}, {Bietenholz}, {van Velzen}, {Anderson}, {Berger}, {Cendes},
  {Chornock}, {Coppejans}, {Eftekhari}, {Gezari}, {Laskar}, {Ramirez-Ruiz}, \&
  {Saxton}}]{Goodwin..2023b}
{Goodwin}, A.~J., {Alexander}, K.~D., {Miller-Jones}, J.~C.~A., {et~al.}
  2023{\natexlab{b}}, \mnras, 522, 5084, \dodoi{10.1093/mnras/stad1258}

\bibitem[{{Goodwin} {et~al.}(2024){Goodwin}, {Anderson}, {Miller-Jones},
  {Malyali}, {Grotova}, {Homan}, {Kawka}, {Krumpe}, {Liu}, \&
  {Rau}}]{Goodwin..2024}
{Goodwin}, A.~J., {Anderson}, G.~E., {Miller-Jones}, J.~C.~A., {et~al.} 2024,
  \mnras, 528, 7123, \dodoi{10.1093/mnras/stae362}

\bibitem[{{Granot} \& {Sari}(2002)}]{Granot2002}
{Granot}, J., \& {Sari}, R. 2002, \apj, 568, 820, \dodoi{10.1086/338966}

\bibitem[{{Hammerstein} {et~al.}(2023){Hammerstein}, {van Velzen}, {Gezari},
  {Cenko}, {Yao}, {Ward}, {Frederick}, {Villanueva}, {Somalwar}, {Graham},
  {Kulkarni}, {Stern}, {Andreoni}, {Bellm}, {Dekany}, {Dhawan}, {Drake},
  {Fremling}, {Gatkine}, {Groom}, {Ho}, {Kasliwal}, {Karambelkar}, {Kool},
  {Masci}, {Medford}, {Perley}, {Purdum}, {van Roestel}, {Sharma}, {Sollerman},
  {Taggart}, \& {Yan}}]{Hammerstein..2023}
{Hammerstein}, E., {van Velzen}, S., {Gezari}, S., {et~al.} 2023, \apj, 942, 9,
  \dodoi{10.3847/1538-4357/aca283}

\bibitem[{{Horesh} {et~al.}(2021){Horesh}, {Cenko}, \&
  {Arcavi}}]{Horesh..2021a}
{Horesh}, A., {Cenko}, S.~B., \& {Arcavi}, I. 2021, Nature Astronomy, 5, 491,
  \dodoi{10.1038/s41550-021-01300-8}

\bibitem[{{Kale} \& {Ishwara-Chandra}(2021)}]{Kare..2021}
{Kale}, R., \& {Ishwara-Chandra}, C.~H. 2021, Experimental Astronomy, 51, 95,
  \dodoi{10.1007/s10686-020-09677-6}

\bibitem[{{Kellermann} {et~al.}(1969){Kellermann}, {Pauliny-Toth}, \&
  {Williams}}]{Kellermann..1969}
{Kellermann}, K.~I., {Pauliny-Toth}, I.~I.~K., \& {Williams}, P.~J.~S. 1969,
  \apj, 157, 1, \dodoi{10.1086/150046}

\bibitem[{{Lei} {et~al.}(2016){Lei}, {Yuan}, {Zhang}, \& {Wang}}]{Lei..2016}
{Lei}, W.-H., {Yuan}, Q., {Zhang}, B., \& {Wang}, D. 2016, \apj, 816, 20,
  \dodoi{10.3847/0004-637X/816/1/20}

\bibitem[{{Lei} {et~al.}(2024){Lei}, {Wu}, {Li}, {Li}, {Lei}, {Fan}, {Wu},
  {Wang}, \& {Yang}}]{Lei..2024}
{Lei}, X., {Wu}, Q., {Li}, H., {et~al.} 2024, \apj, 977, 63,
  \dodoi{10.3847/1538-4357/ad8ba5}

\bibitem[{{Lovell} {et~al.}(2003){Lovell}, {Jauncey}, {Bignall},
  {Kedziora-Chudczer}, {Macquart}, {Rickett}, \& {Tzioumis}}]{Lovell..2003}
{Lovell}, J.~E.~J., {Jauncey}, D.~L., {Bignall}, H.~E., {et~al.} 2003, \aj,
  126, 1699, \dodoi{10.1086/378053}

\bibitem[{{Lu} \& {Kumar}(2018)}]{Lu..2018}
{Lu}, W., \& {Kumar}, P. 2018, \apj, 865, 128, \dodoi{10.3847/1538-4357/aad54a}

\bibitem[{{Lu} {et~al.}(2024){Lu}, {Matsumoto}, \& {Matzner}}]{Lu..2024}
{Lu}, W., {Matsumoto}, T., \& {Matzner}, C.~D. 2024, \mnras, 533, 979,
  \dodoi{10.1093/mnras/stae1770}

\bibitem[{{Matsumoto} \& {Piran}(2023)}]{Matsumoto..2023}
{Matsumoto}, T., \& {Piran}, T. 2023, \mnras, 522, 4565,
  \dodoi{10.1093/mnras/stad1269}

\bibitem[{{Matsumoto} \& {Piran}(2024)}]{Matsumoto..2024}
---. 2024, \apj, 971, 49, \dodoi{10.3847/1538-4357/ad58ba}

\bibitem[{{Mattila} {et~al.}(2018){Mattila}, {P{\'e}rez-Torres}, {Efstathiou},
  {Mimica}, {Fraser}, {Kankare}, {Alberdi}, {Aloy}, {Heikkil{\"a}}, {Jonker},
  {Lundqvist}, {Mart{\'\i}-Vidal}, {Meikle}, {Romero-Ca{\~n}izales}, {Smartt},
  {Tsygankov}, {Varenius}, {Alonso-Herrero}, {Bondi}, {Fransson},
  {Herrero-Illana}, {Kangas}, {Kotak}, {Ram{\'\i}rez-Olivencia},
  {V{\"a}is{\"a}nen}, {Beswick}, {Clements}, {Greimel}, {Harmanen},
  {Kotilainen}, {Nandra}, {Reynolds}, {Ryder}, {Walton}, {Wiik}, \&
  {{\"O}stlin}}]{Mattila..2018}
{Mattila}, S., {P{\'e}rez-Torres}, M., {Efstathiou}, A., {et~al.} 2018,
  Science, 361, 482, \dodoi{10.1126/science.aao4669}

\bibitem[{{McConnell} {et~al.}(2020){McConnell}, {Hale}, {Lenc}, {Banfield},
  {Heald}, {Hotan}, {Leung}, {Moss}, {Murphy}, {O'Brien}, {Pritchard}, {Raja},
  {Sadler}, {Stewart}, {Thomson}, {Whiting}, {Allison}, {Amy}, {Anderson},
  {Ball}, {Bannister}, {Bell}, {Bock}, {Bolton}, {Bunton}, {Chippendale},
  {Collier}, {Cooray}, {Cornwell}, {Diamond}, {Edwards}, {Gupta}, {Hayman},
  {Heywood}, {Jackson}, {Koribalski}, {Lee-Waddell}, {McClure-Griffiths}, {Ng},
  {Norris}, {Phillips}, {Reynolds}, {Roxby}, {Schinckel}, {Shields},
  {Tremblay}, {Tzioumis}, {Voronkov}, \& {Westmeier}}]{McConnell..2020}
{McConnell}, D., {Hale}, C.~L., {Lenc}, E., {et~al.} 2020, \pasa, 37, e048,
  \dodoi{10.1017/pasa.2020.41}

\bibitem[{{Moldon}(2021)}]{Moldon2021}
{Moldon}, J. 2021, {eMCP: e-MERLIN CASA pipeline}, Astrophysics Source Code
  Library, record ascl:2109.006

\bibitem[{{Mou} {et~al.}(2022){Mou}, {Wang}, {Wang}, \& {Yang}}]{Mou..2022}
{Mou}, G., {Wang}, T., {Wang}, W., \& {Yang}, J. 2022, \mnras, 510, 3650,
  \dodoi{10.1093/mnras/stab3742}

\bibitem[{{Mou} \& {Wang}(2021)}]{Mou&Wang2021}
{Mou}, G., \& {Wang}, W. 2021, \mnras, 507, 1684,
  \dodoi{10.1093/mnras/stab2261}

\bibitem[{{Mou} {et~al.}(2021){Mou}, {Dou}, {Jiang}, {Wang}, {Guo}, {Wang},
  {Wang}, {Shu}, {He}, {Zhang}, \& {Sun}}]{Mou..2021}
{Mou}, G., {Dou}, L., {Jiang}, N., {et~al.} 2021, \apj, 908, 197,
  \dodoi{10.3847/1538-4357/abd475}

\bibitem[{{Murphy} {et~al.}(2021){Murphy}, {Kaplan}, {Stewart}, {O'Brien},
  {Lenc}, {Pintaldi}, {Pritchard}, {Dobie}, {Fox}, {Leung}, {An}, {Bell},
  {Broderick}, {Chatterjee}, {Dai}, {d'Antonio}, {Doyle}, {Gaensler}, {Heald},
  {Horesh}, {Jones}, {McConnell}, {Moss}, {Raja}, {Ramsay}, {Ryder}, {Sadler},
  {Sivakoff}, {Wang}, {Wang}, {Wheatland}, {Whiting}, {Allison}, {Anderson},
  {Ball}, {Bannister}, {Bock}, {Bolton}, {Bunton}, {Chekkala}, {Chippendale},
  {Cooray}, {Gupta}, {Hayman}, {Jeganathan}, {Koribalski}, {Lee-Waddell},
  {Mahony}, {Marvil}, {McClure-Griffiths}, {Mirtschin}, {Ng}, {Pearce},
  {Phillips}, \& {Voronkov}}]{Murphy..2021}
{Murphy}, T., {Kaplan}, D.~L., {Stewart}, A.~J., {et~al.} 2021, \pasa, 38,
  e054, \dodoi{10.1017/pasa.2021.44}

\bibitem[{{Pasham} {et~al.}(2015){Pasham}, {Cenko}, {Levan}, {Bower}, {Horesh},
  {Brown}, {Dolan}, {Wiersema}, {Filippenko}, {Fruchter}, {Greiner}, {O'Brien},
  {Page}, {Rau}, \& {Tanvir}}]{Pasham..2015}
{Pasham}, D.~R., {Cenko}, S.~B., {Levan}, A.~J., {et~al.} 2015, \apj, 805, 68,
  \dodoi{10.1088/0004-637X/805/1/68}

\bibitem[{{Rau} \& {Cornwell}(2011)}]{Rau..2011}
{Rau}, U., \& {Cornwell}, T.~J. 2011, \aap, 532, A71,
  \dodoi{10.1051/0004-6361/201117104}

\bibitem[{{Readhead}(1994)}]{Readhead1994}
{Readhead}, A. C.~S. 1994, \apj, 426, 51, \dodoi{10.1086/174038}

\bibitem[{{Rees}(1988)}]{Rees1988}
{Rees}, M.~J. 1988, \nat, 333, 523, \dodoi{10.1038/333523a0}

\bibitem[{{Rickett}(2007)}]{Rickett2007}
{Rickett}, B.~J. 2007, Astronomical and Astrophysical Transactions, 26, 429,
  \dodoi{10.1080/10556790701600580}

\bibitem[{{Saxton} {et~al.}(2020){Saxton}, {Komossa}, {Auchettl}, \&
  {Jonker}}]{Saxton..2020}
{Saxton}, R., {Komossa}, S., {Auchettl}, K., \& {Jonker}, P.~G. 2020, \ssr,
  216, 85, \dodoi{10.1007/s11214-020-00708-4}

\bibitem[{{Sfaradi} {et~al.}(2024){Sfaradi}, {Beniamini}, {Horesh}, {Piran},
  {Bright}, {Rhodes}, {Williams}, {Fender}, {Leung}, {Murphy}, \&
  {Green}}]{Sfaradi..2024}
{Sfaradi}, I., {Beniamini}, P., {Horesh}, A., {et~al.} 2024, \mnras, 527, 7672,
  \dodoi{10.1093/mnras/stad3717}

\bibitem[{{Sfaradi} {et~al.}(2025){Sfaradi}, {Margutti}, {Chornock},
  {Alexander}, {Metzger}, {Beniamini}, {Barniol Duran}, {Yao}, {Horesh},
  {Farah}, {Berger}, {Nayana A.}, {Cendes}, {Eftekhari}, {Fender}, {Franz},
  {Green}, {Hammerstein}, {Lu}, {Wiston}, {Bernstein}, {Bright}, {Christy},
  {Cruz}, {DeBoer}, {Golay}, {Goodwin}, {Gurwell}, {Keating}, {Laskar},
  {Miller-Jones}, {Pollak}, {Rao}, {Siemion}, {Sheikh}, {Shoval}, \& {van
  Velzen}}]{Sfaradi2025}
{Sfaradi}, I., {Margutti}, R., {Chornock}, R., {et~al.} 2025, arXiv e-prints,
  arXiv:2508.03807, \dodoi{10.48550/arXiv.2508.03807}

\bibitem[{{Shepherd}(1997)}]{Shepherd1997}
{Shepherd}, M.~C. 1997, in Astronomical Society of the Pacific Conference
  Series, Vol. 125, Astronomical Data Analysis Software and Systems VI, ed.
  G.~{Hunt} \& H.~{Payne}, 77

\bibitem[{{Shu} {et~al.}(2018){Shu}, {Wang}, {Dou}, {Jiang}, {Wang}, \&
  {Wang}}]{Shu..2018}
{Shu}, X.~W., {Wang}, S.~S., {Dou}, L.~M., {et~al.} 2018, \apjl, 857, L16,
  \dodoi{10.3847/2041-8213/aaba17}

\bibitem[{{Sun} {et~al.}(2025){Sun}, {Guo}, {Gu}, {Li}, {Chen},
  {Gonz{\'a}lez-Buitrago}, {Wang}, {Li}, {Feng}, {Xiong}, {Wang}, {Yuan},
  {Jin}, {Zhang}, {Deng}, \& {Zhang}}]{Sun..2025}
{Sun}, J., {Guo}, H., {Gu}, M., {et~al.} 2025, \apj, 982, 150,
  \dodoi{10.3847/1538-4357/adb724}

\bibitem[{{Sun} {et~al.}(2024){Sun}, {Jiang}, {Dou}, {Shu}, {Zhu}, {Dong},
  {Buckley}, {Bradley Cenko}, {Fan}, {Gromadzki}, {Liu}, {Wang}, {Wang},
  {Wang}, {Wu}, {Yang}, {Zhang}, {Zhang}, \& {Zhang}}]{SunLM..2024}
{Sun}, L., {Jiang}, N., {Dou}, L., {et~al.} 2024, \aap, 692, A262,
  \dodoi{10.1051/0004-6361/202452380}

\bibitem[{{Sutherland} \& {Dopita}(1993)}]{Sutherland1993}
{Sutherland}, R.~S., \& {Dopita}, M.~A. 1993, \apjs, 88, 253,
  \dodoi{10.1086/191823}

\bibitem[{{Teboul} \& {Metzger}(2023)}]{Teboul..2023}
{Teboul}, O., \& {Metzger}, B.~D. 2023, \apjl, 957, L9,
  \dodoi{10.3847/2041-8213/ad0037}

\bibitem[{{Ulvestad} {et~al.}(2005){Ulvestad}, {Antonucci}, \&
  {Barvainis}}]{Ulvestad..2005}
{Ulvestad}, J.~S., {Antonucci}, R. R.~J., \& {Barvainis}, R. 2005, \apj, 621,
  123, \dodoi{10.1086/427426}

\bibitem[{{Veres} {et~al.}(2024){Veres}, {Franckowiak}, {van Velzen},
  {Adebahr}, {Taziaux}, {Necker}, {Stein}, {Kier}, {Mueller}, {Bomans},
  {Jordana-Mitjans}, {Kowalski}, {Hammerstein}, {Marci-Boehncke}, {Reusch},
  {Garrappa}, {Rose}, \& {Kashyap Das}}]{Veres..2024}
{Veres}, P.~M., {Franckowiak}, A., {van Velzen}, S., {et~al.} 2024, arXiv
  e-prints, arXiv:2408.17419, \dodoi{10.48550/arXiv.2408.17419}

\bibitem[{{Walker}(1998)}]{Walker1998}
{Walker}, M.~A. 1998, \mnras, 294, 307,
  \dodoi{10.1046/j.1365-8711.1998.01238.x10.1111/j.1365-8711.1998.01238.x}

\bibitem[{{Yang} {et~al.}(2022){Yang}, {Shu}, {Zhang}, {Chandola}, {Liu},
  {Liu}, {Gu}, {Giustini}, {Jiang}, {Li}, {Li}, {Elbaz}, {Juneau}, {Pannella},
  {Sun}, {Tang}, {Wang}, \& {Zhou}}]{Yang..2022}
{Yang}, L., {Shu}, X., {Zhang}, F., {et~al.} 2022, \apj, 935, 115,
  \dodoi{10.3847/1538-4357/ac80ba}

\bibitem[{{Zhang} {et~al.}(2024){Zhang}, {Shu}, {Yang}, {Sun}, {Zhang}, {Wang},
  {Mou}, {Zhang}, {Zhou}, \& {Peng}}]{Zhang..2024}
{Zhang}, F., {Shu}, X., {Yang}, L., {et~al.} 2024, \apjl, 962, L18,
  \dodoi{10.3847/2041-8213/ad1d61}

\bibitem[{{Zhang} {et~al.}(2022){Zhang}, {Shu}, {Sheng}, {Sun}, {Dou}, {Jiang},
  {Wang}, {Hu}, {Wang}, \& {Wang}}]{Zhang2022}
{Zhang}, W.~J., {Shu}, X.~W., {Sheng}, Z.~F., {et~al.} 2022, \aap, 660, A119,
  \dodoi{10.1051/0004-6361/202142253}

\bibitem[{{Zhuang} {et~al.}(2025){Zhuang}, {Shen}, {Mou}, \&
  {Lu}}]{Zhuang..2025}
{Zhuang}, J., {Shen}, R.-F., {Mou}, G., \& {Lu}, W. 2025, \apj, 979, 109,
  \dodoi{10.3847/1538-4357/ad9b98}

\end{thebibliography}

\end{document}